\documentclass[12pt]{article}

\usepackage{latexsym}
\font\of=msbm10 scaled 1200
\font\ofs=msbm10
\def\R{\mbox{\of R}}
\def\C{\mbox{\of C}}
\def\Z{\mbox{\of Z}}
\def\N{\mbox{\of N}}
\def\Rs{\mbox{\ofs R}}
\begin{document}
\begin{center}
{\Large\bf Stability of periodic
travelling shallow-water

\vspace{1ex}
waves determined by Newton's equation}

\vspace{4ex}
{\bf Sevdzhan Hakkaev$^1$, Iliya D. Iliev$^2$  and Kiril Kirchev$^2$ }
\end{center}

\small

$^1$Faculty of Mathematics and Informatics, Shumen University,

9712 Shumen, Bulgaria

\vspace{1ex}
 $^2$Institute of Mathematics and Informatics, Bulgarian Academy of Sciences,

 1113 Sofia, Bulgaria

\vspace{1ex}
E-mail: {\tt shakkaev@fmi.shu-bg.net, iliya@math.bas.bg, kpkirchev@abv.bg}

\vspace{4ex}
\noindent
{\bf Abstract.} We study the existence and stability of
periodic travelling-wave solutions for generalized
Benjamin-Bona-Mahony and Camassa-Holm equations. To prove orbital
stability, we use the abstract results of Grillakis-Shatah-Strauss
and the Floquet theory for periodic eigenvalue problems.

\normalsize
\vspace{2ex}

\vspace{3ex}
\noindent
{\large\bf 1. Introduction.}

\vspace{1ex}
\noindent
  Consider the following equation
  $$ u_t+(a(u))_x-u_{xxt}=\left(
     b'(u){\frac{u_x^2}{2}}+b(u)u_{xx} \right)_x \eqno(1.1)$$
where $a,b: \R \rightarrow \R$ are smooth functions and $a(0)=0$.
In this paper we study the problems of the existence and stability
of periodic travelling-wave solutions $u=\varphi(x-vt)$ for (1.1).
It is easy to see that whatever $a,b$ be, the equation for $\varphi$ has no
dissipative terms. Hence, any travelling-wave solution of (1.1) is determined
from Newton's equation which we will write below in the form
$\varphi'^2=U(\varphi)$. Therefore by using the well-known properties of the
phase portrait of Newton's equation in the $(\varphi,\varphi')$-plane,
one can establish that under fairly broad conditions, (1.1) has at least one
three-parameter family of periodic solutions
$\varphi(y)=\varphi(v,c_1;\varphi_0;y)$ where $c_1$ is a constant
of integration and $\varphi_0 =\min \varphi$ (see Proposition 1).
The parameters $v$, $c_1$ determine the phase portrait whilst $\varphi_0$
serves to fix the periodic orbit within.
Moreover, if $T=T(v,c_1,\varphi_0)$ is the minimal (sometimes called
{\it fundamental}) period of $\varphi$, then $\varphi$ has exactly one local
minimum and one local maximum in $[0,T)$. Therefore $\varphi'$ has just
two zeroes in each semi-open interval of length $T$. By Floquet theory,
this means that $\varphi'$ is either the second or the third eigenfunction
of the periodic eigenvalue problem obtained from the second variation
along $\varphi$ of an appropriate conservative functional $M(u)$.
If the first case occurs, then one can use the abstract result of
Grillakis-Shatah-Strauss (\cite{GrShSt}) to prove orbital stability
whenever $\ddot d(v)=(d^2/dv^2)M(\varphi)$ is positive.

In the periodic case we deal with, it is not always so easy to determine
the sign of $\ddot d(v)$. To overcome this problem, we first establish a
general result (see Proposition 6) expressing $\ddot d(v)$ through
some special line integrals along the energy level orbit $\{H=h\}$ of the
Newtonian function $H(X,Y)=Y^2-U(X)$ which corresponds to $\varphi$.
When $a,b$ are polynomials, these are complete Abelian integrals and one
can apply methods from algebraic geometry (Picard-Fuchs equations, etc.)
to determine the possible values of $v$, $c_1$ and $\varphi_0$ where
$\ddot d(v)$ changes sign. Let us mention that even for $v$ and $c_1$
fixed, the sign of $\ddot d(v)$ might depend on the amplitude of $\varphi$
(ruled by $\varphi_0$) as shown in Proposition 8. In this connection,
we calculate explicitly the main term of $\ddot d(v)$ in the case of
arbitrary small-amplitude periodic solutions $\varphi$ of (1.1), see formula
(7.7). It is shown that the main term depends on the first two isochronous
constants related to the center $(X_0,0)$ into which the orbit
$(\varphi,\varphi')$ shrinks when $\varepsilon=\max\varphi-\min\varphi\to 0$,
and on $X_0$ itself as well.

We apply our results to prove orbital stability for several particular
examples.

\vspace{2ex}
\noindent
{\bf Theorem I.} (The modified BBM equation).
{\it Let $a(u)=2\omega u+\beta u^3$, $b(u)=0$, $\beta>0$ and
$u=\varphi(x-vt)$ where $v>0$, $\varphi(y)=\varphi(v,0;\varphi_0;y)$
be a periodic travelling-wave solution of $(1.1)$ which does not oscillate
around zero. Then $\varphi$ is orbitally stable in any of the cases:}

\vspace{1ex}
(i) $\;3v^2-8\omega^2\geq 0;$

\vspace{1ex}
(ii) $\;3v^2-8\omega^2<0$, $\;2v^2-2\omega v-\omega^2>0$
{\it and the periog of $\varphi$ is sufficiently large.}

\vspace{2ex}
\noindent
{\bf Theorem II.} (The perturbed single-power BBM equation).
{\it Let $a(u)=\beta u^2$, $b(u)=\gamma\beta u$, $\beta>0$,
and let $u=\varphi(x-vt)$ where $v>0$, $\varphi(y)=\varphi(v,0;\varphi_0;y)$
be a periodic travelling-wave solution of $(1.1)$.
Then $\varphi$ is orbitally stable for small $|\gamma|$.}

\vspace{2ex}
\noindent
{\bf Theorem III.} (The perturbed single-power mBBM equation).
{\it Let $a(u)=\beta u^3$, $b(u)=\gamma\beta u^2$, $\beta>0$,
and let $u=\varphi(x-vt)$ where $v>0$, $\varphi(y)=\varphi(v,0;\varphi_0;y)$
be a periodic travelling-wave solution of $(1.1)$ which does not oscillate
around zero. Then $\varphi$ is orbitally stable for small $|\gamma|$.}

\vspace{2ex}
\noindent
{\bf Theorem IV.} (Small-amplitude waves of the perturbed BBM equation.)
{\it Let $a(u)=2\omega u+\frac32u^2$, $b(u)=\gamma g(u)$ and
$u=\varphi(x-vt)$ where $v>0$, $\varphi(y)=\varphi(v,c_1;\varphi_0;y)$
be a periodic travelling-wave solution of $(1.1)$ having a small amplitude.
Then $\varphi$ is orbitally stable for small $|\gamma|$ and
$(\omega/v,c_1/v^2)$ taken in appropriate domain $\Omega\subset\R^2$.}

\vspace{2ex}
\noindent
{\bf Theorem V.} (Small-amplitude waves of the perturbed mBBM equation).
{\it Let $a(u)=2\omega u+\beta u^3$, $b(u)=\gamma g(u)$, $\beta>0$ and
$u=\varphi(x-vt)$ where $v>0$, $\varphi(y)=\varphi(v,0;\varphi_0;y)$
be a periodic travelling-wave solution of $(1.1)$ which has a small
amplitude and does not oscillate around zero. Then $\varphi$ is orbitally
stable for $3v^2-8\omega^2>0$ and small $|\gamma|$.}

\vspace{2ex}
\noindent
We point out that, unlike the other cases, in Theorem IV the constant of
integration $c_1$ is not fixed, therefore we consider the whole family
of small-amplitude waves. The explicit expression of $\Omega$ is given in
the proof.

\vspace{2ex}
Let us mention that for $a(u)=2ku+{\frac{3}{2}}u^2$ and $b(u)=u$,
equation (1.1) becomes the well-known Camassa-Holm equation
 $$ u_t+2ku_x+3uu_x-u_{xxt}=2u_xu_{xx}+uu_{xxx}. \eqno(1.2)$$
Equation (1.2) was derived as a bi-Hamiltonian generalization of
the Korteweg-de Vries equation \cite{FoFu} and later Camassa and
Holm \cite{CaHo} recovered it as a water-wave model. The
Camassa-Holm equation is locally well-posed in $H^s$ for
$s>{\frac{3}{2}}$. Moreover, while some solutions of equation
(1.2) are global, others blow up in finite time (in both the
periodic and non-periodic cases)
\cite{Co1,Co2,CoEs1,CoEs2,CoEs3,LiOl,Yin2}. The solitary waves of
Camassa-Holm equation are smooth in the case $k>0$ and peaked for
$k=0$. Their stability is considered in
\cite{CoMo,CoSt1,CoSt2,HaKi1,HaKi2,Le1,Le2}.

For $a(u)=2ku+{\frac{3}{2}}u^2$ and $b(u)=\gamma u$,
equation (1.1) serves as a model equation for mechanical
vibrations in a compressible elastic rod \cite{Da,DaHu}. Some
problems such as well-posedness and blowing-up in this case were
studied in \cite{Yin1,Yin2,Zh}, and stability of solitary
waves was investigated in \cite{CoSt3,Ka}.

If $b(u)=0$ in (1.1), one obtains the generalized Benjamin-Bona-Mahony (gBBM)
equation for surface waves in a channel \cite{BBM}. All solutions
are global and their solitary waves are stable or unstable
depending on $a(u)$ \cite{IlKi,SoSt}.

For equation (1.1), the well-posedness and stability of
solitary waves in the case $a(u)=2ku+{\frac{p+2}{2}}u^{p+1}$ and
$b(u)=u^p$ are studied in \cite{HaKi1}. For $k=0$, equation
(1.1) admits peaked solitary wave solutions, which are stable
(see \cite{HaKi2}).

The existence and stability of periodic travelling waves for
nonlinear evolution equations has received little attention.
Recently Angulo, Bona and Scialom in \cite{ABS} developed a
 complete theory of the stability of cnoidal waves for KdV equation.
 The solution $u(x,t)=\varphi_c(x-ct)$ of KdV satisfies the equation
   $$\varphi_c^{''}+{\textstyle\frac12}\varphi_c^2-c\varphi_c=A_{\varphi_c},$$
 where $A_{\varphi_c}$ is an integration constant. An explicit form for
 $\varphi_c$ in the periodic case is
  $$\varphi_c(\xi)=\beta_2+(\beta_3-\beta_2)cn^2\left(
  \sqrt{\frac{\beta_3-\beta_1} {12}} \xi ; k \right), $$
  where $cn$ is the Jacobi elliptic function and
  the following relations take place:
   $$
      \beta_1<\beta_2<\beta_3,\quad
      k^2={\frac{\beta_3-\beta_2}{\beta_3-\beta_1}},\quad
      \beta_1+\beta_2+\beta_3=3c, \quad
      A_{\varphi_c}=-{\textstyle\frac16}\Sigma_{i<j}{\beta_i
      \beta_j}.$$
Since $cn(u+2K)=-cn(u)$ where $K=K(k)=
\int_{0}^{1}{{\frac{dt}{\sqrt{(1-t^2)(1-k^2t^2)}}}}$ is the complete
elliptic integral of the first kind, then $\varphi_c$ has the
fundamental (i.e. minimal) period $T_{\varphi_c}$ given by
   $$T_{\varphi_c}={\frac{4\sqrt{3}}{\sqrt{\beta_3-\beta_1}}}K(k).$$
 Moreover, $T_{\varphi_c}$ depends on the speed $c$ and satisfies
 the inequality
   $$T_{\varphi_c}^2>{\frac{(2\pi)^2}{\sqrt{c^2+2A_{\varphi_c}}}}$$
  Using the abstract results of
 Grillakis, Shatah and Strauss (adapted to the periodic context),
 the authors proved in \cite{ABS} that the cnoidal waves with mean
 value zero are orbitally stable.

 Other new explicit formulas for the periodic
 travelling waves based on the Jacobi elliptic function of
 type dnoidal, together with their stability, have been obtained
 by Angulo \cite{An,An2} for the nonlinear Schr\"odinger (NLS) equation
   $iu_t+u_{xx}+|u|^2u=0,$
 modified KdV equation $u_t+3u^2u_x+u_{xxx}=0$
 and Hirota-Satsuma system
  $$
    \left\{
        \begin{array}{ll}
          u_t-a(u_{xxx}+6uu_x)=2bvv_x\\
          v_t+v_{xxx}+3uv_x=0.
        \end{array} \right. $$
  For NLS the solutions are of the form $u(x,t)=e^{iwt}\varphi_{w}(x)$
  where $\varphi_w(x)$ is a periodic smooth function with period
  $L>0$. The solution $\varphi_w$ is of the form
    $$\varphi_w(x)=\eta_1 dn\left({\frac{\eta_1}{\sqrt{c}}}x;
    k\right) $$
  where $\eta_1$ and the modulus $k$ depend smoothly on $w$.
  The orbit $\Omega_{\varphi_w}$,
    $$\Omega_{\varphi_w}=\{ e^{i\theta }\varphi_w(\cdot+y), \; \; (y,
    \theta )\in \R \times [0,2\pi) \}$$
  generated by the dnoidal wave $\varphi_w$ is stable by
  perturbation of periodic function with period $L$ and
  nonlinearly unstable by perturbation of periodic function
  with period $2L$.
   In all these works it was necessary to use an elaborated
   spectral theory for the periodic eigenvalue problem,
    $$
     \left\{
          \begin{array}{ll}
             {\frac{d^2}{dx^2}}\Psi+[\rho-n(n+1)k^2sn^2(x;k)]
             \Psi=0,\\
             \Psi(0)=\Psi(2K(k)), \; \; \Psi'(0)=\Psi'(2K(k)),
          \end{array} \right. \eqno{(1.3)}
   $$
with specific values of $n\in \N$. We will also make use of systems
similar to (1.3).

The present paper is organized as follows. In section 2 we formulate and sketch
the proof of a local well-posedness result for the equation (1.1) in periodic
$H^s$ spaces. In Section 3 we prove the existence of periodic travelling
waves of a given (admissible) period and study their properties.
In Section 4 we prove the orbital stability result under some hypotheses
(see Assumption 1). In Section 5 several particular examples are considered.
For most of them, we determine $\varphi$ explicitly and show that
$\varphi'$ is the second eigenfunction of the respective periodic eigenvalue
problem. We also determine the sign of $\ddot d(v)$ to outline the cases
satisfying Assumption 1. In Section 6  perturbation theory is applied
to prove the orbital stability of periodic travelling waves for generalized
Benjamin-Bona-Mahony and generalized Camassa-Holm equations (the case of
small $b$ in the right-hand side of (1.1)). In Section 7 we study the
small-amplitude periodic travelling waves of (1.1) and determine the sign of
$\ddot d(v)$ for them.

\vspace{2ex}
\noindent
{\large\bf 2. Local well-posedness.}

\vspace{1ex}
\noindent
In this section, we discuss the local well-posedness of the Cauchy problem
for equation (1.1). We begin by introducing some notation and by recalling
related definitions we shall use throughout the paper.

Let ${\cal P}=C_{per}^{\infty}$ denote the
collection of all functions which are $C^{\infty}$ and periodic
with a period $T>0$. The topological dual of ${\cal P}$ will be
denoted by ${\cal P'}$. If $\Psi \in {\cal P}'$ then we denote by
$\Psi(f)=\langle \Psi, f \rangle$ the value of $\Psi$ at $f$.
Define the functions $\Theta_k(x)=\exp(2\pi ikx/T), k\in \Z$. The
Fourier transform of $\Psi \in {\cal P'}$ is the function
$\widehat{\Psi}: \Z \rightarrow \C$ defined by
$\widehat{\Psi}(k)={\frac{1}{T}}\langle \Psi, \Theta_{-k}\rangle
$. If $\Psi$ is a periodic function with a period $T$, we have
  $$ \widehat{\Psi}(k)=\frac{1}{T}\int_0^T\Psi(x)\exp(-2\pi ikx/T)dx.$$
For $s\in \R$, the Sobolev space $H^s([0,T])$ is the set of all
$f\in {\cal P}'$ such that
$$ ||f||_s^2=T\sum\limits_{k=-\infty}^\infty(1+|k|^2)^s|\widehat{f}(k)|^2
  < \infty.$$
Certainly,
  $H^s([0,T])$ is a Hilbert space with respect to the inner product
    $$(f,g)_s=T\sum\limits_{k=-\infty}^\infty (1+|k|^2)^s\widehat{f}(k)
    \overline{\widehat{g}(k)}.$$
  Since $H^s([0,T])\subset L^2([0,T])$ for every $s\geq 0$,
  we obtain via Plancherel identity that for every $n\in \N$,
    $$ ||f||_n^2=\sum\limits_{j=0}^n||f^{(j)}||^2. $$
  where $f^{(j)}$ represents the  $j$th derivative of $f$ taken in
  the sense of ${\cal P}'$. Moreover, Sobolev's lemma states that
  if $s>l+\frac12$, then $H^2([0,T])\hookrightarrow C_{per}^l$ where
$$C_{per}^l=\{ f\in C^l:\; f^{(j)}\;\;\mbox{\rm is periodic with a period}
\; T\;\mbox{\rm for}\; j=0,...,l\}.$$

 One can prove the following result about equation (1.1).

\vspace{2ex}
\noindent
{\bf Theorem 1.} {\it Assume that $a,b \in C^{m+3}(\R)$,
  $m\geq 2$. Given $u_0 \in H^s$, ${\frac{3}{2}}<s<m$,
  there exists a maximal $t_0>0$ and a unique
  solution $u(x,t)$ to $(1.1)$ such that
    $$u \in C([0,t_0), H^s)\cap C^1([0, t_0), H^{s-1}). $$
  Moreover, the solution depends continuously on the initial data. }

\vspace{2ex}
    \noindent
{\bf Proof.} Take $u \in H^s$ and let
   $$ A(u)=b(u)\partial_x, \;\; f(u)=(1-\partial_x^2)^{-1}[b(u)u_x-
   \textstyle{ \partial_x(\frac12b^{'}(u)u_x^2+a(u))}].$$
 Using the above notations, one can rewrite  equation (1.1) in the
 following form:
     $$ u_t+A(u)u=f(u).$$
In a similar way   as in Theorem 2.2
   in \cite{HaKi2} (dealing with the non-periodic case), we have

\vspace{1ex}
\noindent
   (1) $A(u)$ is quasi-m-accretive, uniformly on the bounded sets in
   $H^{s-1}$. Moreover, $A(u) \in L(H^s, H^{s-1})$ (where $L(X,Y)$ is
   the space of all linear bounded operators from $X$ to $Y$, $L(X)=L(X,X)$)
   and
     $$||(A(u_1)-A(u_2))u_3||_{s-1}\leq \mu_1||u_1-u_2||_{s-1}||u_3||_s. $$

\vspace{1ex}
\noindent
   (2) Define  $\Lambda=(1-\partial_x^2)^{1/2}$,
   $B(u)=[\Lambda, b(u)\partial_x ]\Lambda^{-1}$ for
   $u\in H^s$, where $[\Lambda,\mbox{\rm M}]$ denotes the commutator of
   $\Lambda$ and $\mbox{\rm M}$. Then $B(u)\in L(H^{s-1})$ and
      $$ ||(B(u_1)-B(u_2))u_3||_{s-1}\leq
      \mu_2||u_1-u_2||_s||u_3||_{s-1}, \; \; u_1, u_2\in H^s, \;
      \; u_3\in H^{s-1}.$$

\vspace{1ex}
\noindent
   (3) $f(u)$ is bounded on the bounded sets in $H^s$ and satisfies
$$\begin{array}{l}
     ||f(u_1)-f(u_2)||_s \leq \mu_3||u_1-u_2||_s, \; \; u_1,
     u_2\in H^s,\\[2mm]
     ||f(u_1)-f(u_2)||_{s-1} \leq \mu_4||u_1-u_2||_{s-1}, \; \; u_1,
     u_2\in H^{s-1}.\end{array}$$
    Applying Kato's theory for abstract quasilinear evolution
    equations \cite{Kato}, we obtain the local well-posedness of
    the equation (1.1) in $H^s$, for ${\frac{3}{2}}<s<m$. The
    solution $u(x,t)$ belongs to $C([0,t_0), H^s)\cap C^1([0, t_0),
    H^{s-1})$.

\vspace{2ex}
\noindent
{\large\bf 3. Periodic travelling-wave solutions.}

\vspace{1ex}
\noindent
We are looking for a travelling-wave solution of (1.1) of the form
$u(x,t)=\varphi(x-vt)$. We assume that $\varphi$ is smooth and bounded in
$\R$. The following two cases appear:

\vspace{1ex}
(i) $\varphi'\neq 0$ in $\R$ and $\varphi_-<\varphi<\varphi_+$ (corresponding
to kink-wave solution);

\vspace{1ex}
(ii) $\varphi'(\xi)=0$  for some $\xi\in\R$. Denote $\varphi_0=\varphi(\xi)$,
$\varphi_2=\varphi''(\xi)$.

\vspace{2ex}
\noindent
Below we will deal with the second case. Replacing in (1.1) we get
$$-v\varphi'+(a(\varphi))'+v\varphi'''=\left(b'(\varphi)
\frac{\varphi'^2}{2}+b(\varphi)\varphi''\right)'.\eqno(3.1)$$
By integrating (3.1) twice, one obtains
$$-v\varphi+a(\varphi)+v\varphi''=b'(\varphi)\frac{\varphi'^2}{2}
+b(\varphi)\varphi''+c_1,\eqno(3.2)$$
$$[v-b(\varphi)]\frac{\varphi'^2}{2}=
c_2+c_1\varphi+\frac{v}{2}\varphi^2-A(\varphi),\qquad
A(\varphi)=\int_0^\varphi a(s)ds,\eqno(3.3)$$
with some constants $c_1, c_2$. In case (ii), one has respectively
$$\begin{array}{l}
c_1=a(\varphi_0)-v\varphi_0+(v-b(\varphi_0))\varphi_2,\\
c_2=A(\varphi_0)+\frac12v\varphi_0^2-\varphi_0 a(\varphi_0)-
(v-b(\varphi_0))\varphi_0\varphi_2=A(\varphi_0)-\frac12v\varphi_0^2-c_1\varphi_0.\end{array} $$
Next we are going to look for periodic travelling-wave solutions $\varphi$.
Consider in the plane $(X,Y)=(\varphi,\varphi')$ the Hamiltonian system
$$\begin{array}{l}
\dot{X}=[v-b(X)]Y=H_Y,\\
\dot{Y}=\frac12b'(X)Y^2-a(X)+vX+c_1=-H_X,\end{array}\eqno(3.4)$$
with a Hamiltonian function
$$H(X,Y)=[v-b(X)]\frac{Y^2}{2}+A(X)-\frac{v}{2}X^2-c_1X.$$
Then (3.3) becomes $H(\varphi,\varphi')=c_2$ and the curve
$s\to (\varphi(s-s_0),\varphi'(s-s_0))$ determined by (3.3) lies on the energy
level $H=c_2$ of the Hamiltonian $H(X,Y)$. Within the analytical class,
system (3.4) has periodic solutions if and only if it has a center.
Each center is surrounded by a continuous band of periodic trajectories
(called {\it period annulus}) which terminates at a certain separatrix contour
on the Poincar\'e sphere. As far as the straight lines $X=X_0$ where
$v-b(X_0)=0$ are (unions of) trajectories, the critical points of center type
of (3.4) are given by the critical points on $Y=0$ having a negative Hessian.
These are the points $(X_0,0)$ where:
$$c_1+vX_0-a(X_0)=0,\;\;\;  [v-b(X_0)][v-a'(X_0)]<0.\eqno(3.5)$$
(For simplicity, we will not consider here the case of a degenerate center
when the Hessian becomes zero.)

The above considerations lead us to the following statement.

\vspace{2ex}
\noindent
{\bf Proposition 1.} {\it Let $c_1$ and $v$ be constants such that
conditions $(3.5)$ are satisfied for some $X_0\in\R$. Then there is an open
interval $\Delta$ containing $X_0$ such that:}

\vspace{1ex}
(i) {\it For any $\varphi_0\in \Delta$, $\varphi_0<X_0$, the solution of
$(1.2)$
satisfying
$$\varphi(\xi)=\varphi_0,\quad  \varphi'(\xi)=0,\quad
\varphi''(\xi)=\frac{c_1+v\varphi_0-a(\varphi_0)}{v-b(\varphi_0)},$$
is periodic.}

\vspace{1ex}
(ii) {\it If $\varphi_1\in\Delta$, $\varphi_1>X_0$  is the nearest to $X_0$
solution of $H(X,0)=H(\varphi_0,0)$, then $\varphi_0\leq\varphi\leq\varphi_1$.}

\vspace{1ex}
(iii) {\it If $T$ is the minimal period of $\varphi$, then in each
interval $[s,s+T)$ the function $\varphi$ has just one minimum and one
maximum $($equal to $\varphi_0$ and $\varphi_1$, respectively$)$ and is
strictly monotone elsewhere.}

\vspace{2ex}
\noindent
{\bf Proof.} Part (i) follows from the analysis already done above.
Parts (ii) and (iii) follow from the fact that $H(X,Y)$ is symmetric
with respect to $Y$ and because for any fixed $\bar{X}$ the equation
$H(\bar{X},Y)=const$ has just two solutions $\bar{Y}, -\bar{Y}$.

\vspace{2ex}
\noindent
{\bf Example.} Assume that $a$, $b$ are polynomials and $deg\, a=2$,
$deg\, b\leq 1$. For $c_1$, $v$ properly chosen, the quadratic equation in
(3.5) will have two distinct real roots $X_1<X_2$. It is easy to see that if
$v-b(X)\neq 0$ in $[X_1, X_2]$, then (3.5) holds for just one of these roots.
If $a''b'<0$ and $v-b(X)$ vanishes at $X_1$, then (3.5) holds for $X_2$, and
vice versa. Finally, if $a''b'<0$ and $v-b(X)$ has a root within $(X_1, X_2)$,
then both $X_1$ and $X_2$ satisfy (3.5). To summarize: the phase portraits of
all quadratic systems (3.4) (including the exceptional case $deg\,a\leq 1$,
$\deg\,b=1$) satisfying (3.5), that is having a center, are divided into 7
topologically different classes, see e.g. \cite{7}.

\vspace{2ex}
\noindent
{\bf Remark 1.} Below, we shall denote
$\Delta^-=\Delta\cap\{(-\infty,X_0)\}$, $\Delta^+=\Delta\cap\{(X_0,\infty)\}$.
It is possible that (3.3) also has periodic solutions for
initial values $\varphi_0$ far from $X_0$. To study them, it is needed to
specify the functions $a$ and $b$ in (1.1).

\vspace{2ex}
\noindent
Let us denote
$$ U(s)=\frac{2c_2+2c_1s+vs^2-2A(s)}{v-b(s)}=\frac{2A(\varphi_0)-v\varphi_0^2-2c_1\varphi_0+2c_1s+vs^2-2A(s)}{v-b(s)}.$$
Then for $\varphi_0\leq\varphi\leq \varphi_1$ one can rewrite (3.3) as
$\varphi'(\sigma)=\sqrt{U(\varphi(\sigma))}$.  Integrating the equation
along the interval $[\xi,s]\subset [\xi,\xi+T/2]$ yields an implicit
formula for the value of $\varphi(s)$:
$$\int_{\varphi_0}^{\varphi(s)}\frac{d\sigma}{\sqrt{U(\sigma)}}=s-\xi,
\quad  s\in [\xi,\xi+T/2]. \eqno (3.6)$$
For $s\in [\xi+T/2,\xi+T]$ one has  $\varphi(s)=\varphi(T+2\xi-s)$.
We recall that the period function $T$ of a Hamiltonian flow
generated by $H_0\equiv \frac12Y^2-\frac12U(X)=0$ is determined from
$$T=\int_0^Tdt=\oint_{H_0=0}\frac{dX}{Y}=
2\int_{\varphi_0}^{\varphi_1}\frac{dX}{\sqrt{U(X)}} \eqno(3.7)$$
This is in fact the derivative (with respect to the energy level)
of the area surrounded by the periodic trajectory through the
point $(\varphi_0,0)$ in the $(X,Y)=(\varphi, \varphi')$-plane.

\vspace{1ex}
\noindent
Consider the continuous family of periodical travelling-wave solutions
$\{u=\varphi(x-vt)\}$ of (1.1) and (3.3) going through the points
$(\varphi,\varphi')=(\varphi_0,0)$ where $\varphi_0\in\Delta^-$. For any
$\varphi_0\in\Delta^-$, denote by $T=T(\varphi_0)$ the corresponding period.
One can see (e.g. by using formula (3.7) above) that the period function
$\varphi_0\to T(\varphi_0)$ is smooth. To check this, it suffices to
perform a change of the variable
$$X=\frac{\varphi_1-\varphi_0}{2}s+\frac{\varphi_1+\varphi_0}{2}\eqno(3.8)$$
in the integral (3.7) and use that
$$U(\varphi_0)=U(\varphi_1)=0.\eqno(3.9)$$
Also, it is not difficult to verify (see Section 7) that
$$T(\varphi_0)\to T_0=2\pi\sqrt\frac{v-b(X_0)}{a'(X_0)-v}\qquad
\mbox{\rm as}\qquad \varphi_0\uparrow X_0.$$
Conversely, taking $v$, $c_1$ to satisfy the conditions of Proposition 1
and fixing $T$ in a proper interval, one can determine $\varphi_0$
and $\varphi_1$ as smooth functions of $v$, $c_1$ so that the periodic
solution $\varphi$ given by (3.6) will have a period $T$. The condition
for this is the monotonicity of the period.

\vspace{2ex}
\noindent
{\bf Definition 1.} We say that the period $T=T(\varphi_0)$ is {\it not
critical} provided that $T'(\varphi_0)\neq 0$.

\vspace{2ex}
\noindent
If the period $T(\varphi_0)$ is not critical for any $\varphi_0\in\Delta_-$,
then the period function is strictly monotone along the period annulus and
its range is an open interval $I$ having $T_0$ as an endpoint.

\vspace{2ex}
\noindent
{\bf Proposition 2.} {\it Let $\cal A$ be a period annulus of $(3.4)$
which surrounds a nondegenerate center and has a monotone period function.
Then for any $T\in I$ there is a unique $\varphi_0\in\Delta_-$ satisfying
$T(\varphi_0)=T$. Its derivative $\dot \varphi_0$ with respect to $v$ is
determined from}
$$\dot \varphi_0 [c_1+v\varphi_0-a(\varphi_0)]\frac{d}{dh}\oint_{H=h}
\frac{dx}{y}=\frac{d}{dh}\oint_{H=h} \frac{(x^2+y^2-\varphi_0^2)dx}{2y}
\eqno(3.10)$$
{\it where}
$$h=A(\varphi_0)-\frac{v}{2}\varphi_0^2-c_1\varphi_0.\eqno(3.11)$$

\vspace{2ex}
\noindent
{\bf Proof.} We use the implicit function theorem (IFT) and the
Gelfand-Leray form (see \cite{8}, Chapter 3). Let $I$ be the range of the
period function along $\cal A$. For $T\in I$ and $\varphi_0\in\Delta_-$ denote
$$G(v,c_1,\varphi_0)=\bar G(v,c_1,h)=T(\varphi_0)-T=\oint_{H=h}\frac{dx}{y}-T.$$
Then
$$0\neq T'(\varphi_0)= \frac{\partial G}{\partial \varphi_0}=
\frac{dh}{d\varphi_0}\frac{\partial \bar G}{\partial h}=
[a(\varphi_0)-v\varphi_0-c_1]\frac{d}{dh}\oint_{H=h}\frac{dx}{y}$$
and the IFT works. Hence
$$\bar G=\frac{\partial \bar G}{\partial h}\dot h+
\frac{\partial \bar G}{\partial v}=0.\eqno(3.12)$$
From $H(x,y)=h$ we obtain the covariant derivatives
$$(v-b(x))y\frac{dy}{dh}=1, \quad \frac12(y^2-x^2)+(v-b(x))y\dot y=0.
\eqno(3.13)$$
Therefore, by using Gelfand-Leray form to calculate the derivatives, we get
$$\bar G(v,c_1,h)=\frac{d}{dh} \oint_{H=h}(v-b(x))ydx-T,\quad
\frac{\partial \bar G}{\partial h}=\frac{d}{dh}\oint_{H=h}\frac{dx}{y},
\eqno(3.14)$$
$$\frac{\partial \bar G}{\partial v}= \frac{d}{dh}
\oint_{H=h}[y+(v-b(x))\dot y]dx= \frac{d}{dh} \oint_{H=h}
\frac{(x^2+y^2)dx}{2y}.\eqno(3.15)$$
Hence, by (3.12), (3.14) and (3.15) we obtain
$$\dot h \frac{d}{dh} \oint_{H=h}\frac{dx}{y}+
 \frac{d}{dh} \oint_{H=h} \frac{(x^2+y^2)dx}{2y}=0.\eqno(3.16)$$
Finally, from (3.11) one obtains
$\dot h=-\frac12\varphi_0^2+[a(\varphi_0)-v\varphi_0-c_1]\dot\varphi_0$.
Together with (3.16), this implies (3.10). $\Box$

\vspace{2ex}
\noindent
{\bf Remark 2.} Obviously, one can formulate a local variant of Proposition 2
concerning a given noncritical period $T(\varphi_0)$ only. As far as
$T'(\varphi_0)\neq 0$, the same proof clearly goes and no restrictions
concerning the period annulus $\cal A$ are needed.

\vspace{2ex}
\noindent
The perturbation result we establish below will be needed in Section 6.
Instead of (1.1), consider now a small perturbation of the generalized BBM
equation
$$u_t+(a(u))_x-u_{xxt}=\gamma\left(b'(u)\frac{u_x^2}{2}+b(u)u_{xx}\right)_x,
\quad |\gamma|<\!\!<1\eqno(1.1_\gamma)$$ and let
$\{\varphi^\gamma(x-vt)\}$ be the family of corresponding periodic
travelling-wave solutions going through points
$(\varphi_0^\gamma,0)$ in the $(\varphi,\varphi')$-plane. Denote the
related periods by $T(\varphi_0^\gamma)$.

\vspace{2ex}
\noindent
{\bf Proposition 3.} {\it Assume that the period $T=T(\varphi_0^0)$
related to the solution $\varphi^0(x-vt)$ of $(1.1_0)$ is not critical. Then:}

\vspace{1ex} (i) {\it There is a smooth function
$\gamma\to \varphi_0(\gamma)$ defined for small $|\gamma|$ and
satisfying $\varphi_0(0)=\varphi_0^0$, such that the travelling
wave solution $\varphi^\gamma(x-vt)$ of $(1.1_\gamma)$ going
through the point $(\varphi_0(\gamma),0)$ is periodical and has a
$($minimal$\,)$ period $T$.}

\vspace{1ex}
(ii) $\max_{[0,T]}|\varphi^\gamma-\varphi^0|=O(\gamma).$

\vspace{2ex}
\noindent
{\bf Proof.} (i). The proof relies on the implicit function theorem.
Consider system (3.4) with $b$ replaced by $\gamma b$. Since for
$\varphi_0\in\Delta^-$ the point $(\varphi_0,0)$ is not critical
for (3.4), IFT yields that there is a smooth function
$\varphi_1=\varphi_1(\varphi_0,\gamma)$ determined from
$H(\varphi_1,0)=H(\varphi_0,0)$,
which takes values in $\Delta^+$ and $\varphi_0\leq\varphi^\gamma\leq\varphi_1$.
Introducing a new variable (3.8) in (3.7) and making use of (3.9),
we can rewrite (3.7) in the form $G(\gamma, \varphi_0)=0$ where
$$G(\gamma, \varphi_0)=2\int_{-1}^1\frac{ds}{\sqrt{(1-s^2)U_1(\gamma, \varphi_0,s)}}-T,$$
with $U_1$ a smooth positive function. We further have
$$G(0,\varphi_0^0)=0,\qquad \left.\frac{dG(\gamma, \varphi_0)}{d\varphi_0}\right|_{\gamma=0}=
T'(\varphi_0^0)\neq 0,$$
therefore (i) follows, again by the IFT.

(ii) Consider system (3.4) with $b$ replaced by $\gamma b$ and initial data
$X(\xi)=\varphi_0(\gamma)$, $Y(\xi)=0$. By the uniqueness and the
smooth dependence theorems, the solution $\varphi^\gamma$
is smooth and therefore a uniformly Lipschitz continuous function with
respect to $\gamma$ for $|\gamma|$ small, on each interval $[\xi,\xi+T]$. $\Box$

\vspace{2ex}
\noindent
In support to the hypothesis of Proposition 3, we state the following
(in fact, known) result about the period $T=T(\varphi_0)$.

\vspace{2ex}
\noindent
{\bf Proposition 4.} (i) {\it There is no critical period provided
that $a(u)$ in $(1,1_\gamma)$ is a polynomial of degree 2}.
(ii). {\it There is at most one critical period, if $a(u)$ is a
polynomial of degree 3.}

\vspace{2ex}
\noindent
{\bf Proof.} Consider the period function related to the
Hamiltonian system (3.4) (with $b(X)=0$) as a function $T(h)$ of the energy
level $H=h$. It is well known that $T'(h)\neq 0$ if deg $a=2$ \cite{3}
and that $T'(h)$ has at most one zero if deg $a=3$ \cite{4}.
As
$$\frac{d}{d\varphi_0}=[a(\varphi_0)-v\varphi_0-c_1]\frac{d}{dh}$$
and the coefficient is not zero for $\varphi_0\in\Delta^-$
according to (3.5), the claim follows. $\Box$

\vspace{2ex}
\noindent
{\large\bf 4. Conservation laws and conditional stability.}

\vspace{1ex}
\noindent
We now turn to our main stability problem.
Take $v\in\R$ and denote by $\varphi_v$ the
periodic travelling-wave solution $u=\varphi(x-vt)$ of (1.1).

\vspace{2ex}
\noindent
   {\bf Definition 2.} The periodic travelling-wave solution $\varphi_v$ of
   (1.1) is said to be {\it stable}, if for every $\varepsilon >0$
   there is $\delta >0$ such that if $u\in C([0,t_0),H^1)$ is a solution of
   (1.1) satisfying $||u(0)-\varphi_v||_1<\delta $, then
     $$\inf_{r \in \Rs}||u(\cdot,
     t)-\varphi(\cdot-r)||_1<\varepsilon \quad
   \mbox{\rm for every}\; t\in [0,t_0). $$

\vspace{2ex}
\noindent
  For $\varepsilon>0$, consider a neighborhood ${\cal U}_{\varepsilon}$
  in the set of all translations of $\varphi_v$ defined by
  $${\cal U}_{\varepsilon}=\{ u\in H^1([0,T]) : \inf\limits_{r\in
  \Rs}||u-\varphi_v(\cdot -r)||_1<\varepsilon \}.$$

\vspace{2ex}
\noindent
{\bf Lemma 1.} {\it There exists $\varepsilon>0$ and a $C^1$-map
$\alpha : {\cal U}_{\varepsilon} \rightarrow \R/T$ such that for all $u\in
{\cal U}_{\varepsilon}$, one holds}
$\langle u(\cdot +\alpha(u)),\varphi'_v \rangle=0.$

\vspace{2ex}
\noindent
{\bf Proof.} Standard (for details, see Lemma 4.1 in \cite{777}).

\vspace{1ex}
As usual, an important role in our construction will be played by some
functionals which are invariant with respect to time $t$. Equation (1.1)
possesses the following conservation laws:
$$\begin{array}{l}
\displaystyle E(u)=-\int_0^T\left[A(u)+\frac{b(u)}{2}u_x^2\right]dx,\\[5mm]
\displaystyle Q(u)=\frac12\int_0^T(u^2+u_x^2)dx, \\[5mm]
\displaystyle V(u)=\int_0^Tudx \end{array}$$
where $T$ is the
minimal period of the solution $u(x,t)$ and $A'(u)=a(u)$. Let us
denote for short $M=E+vQ$. In terms of $E$, $Q$, and $V$ equation
(3.2) with $\varphi=\varphi_v$ reads
$$M'(\varphi_v)+c_1=E'(\varphi_v)+vQ'(\varphi_v)+c_1V'(\varphi_v)=0.\eqno(4.1)$$
Let
$$ d(v)=M(\varphi_v). \eqno(4.2)$$
Then differentiating (4.2) with respect to $v$ we obtain
$$\begin{array}{l}
\dot d(v)=Q(\varphi_v),\\[2mm]
 {\displaystyle \ddot d(v)=\langle Q'(\varphi_v), \dot\varphi_v \rangle=
 \frac{d}{dv}\left(\frac12\int_0^T(\varphi_v^2+{\varphi'_v}^2)dx
  \right).} \end{array}\eqno(4.3) $$
Consider in $L^2[0,T]$ the operator ${\cal H}_v$ defined by the formal
differential expression
  $${\cal H}_v=M''(\varphi_v)= (b(\varphi_v)-v)\partial_x^2+
b'(\varphi_v)\varphi'_v\partial_x+v-a'(\varphi_v)
+{\textstyle\frac12}b''(\varphi_v)\varphi'^2_v+b'(\varphi_v)\varphi''_v.
\eqno(4.4)$$
As ${\cal H}_v\varphi'_v=0$, zero is in the spectrum of ${\cal H}_v$.
We make the following assumption concerning ${\cal H}_v$ and $\ddot d(v)$:

\vspace{2ex}
\noindent
{\bf Assumption 1.} (i) {\it The operator ${\cal H}_v$ has a unique negative
eigenvalue, a simple eigenvalue $0$ and the rest of its spectrum is positive.}
(ii) $\ddot d(v)>0$.

\vspace{2ex}
\noindent
{\bf Lemma 2.} {\it If Assumption 1 holds and $y$ satisfies
$\langle Q'(\varphi_v), y \rangle = \langle
\varphi_v', y \rangle =0$, then $\langle{\cal H}_v y, y \rangle >0$.}

\vspace{2ex}
\noindent
{\bf Proof.} Differentiating (4.1) with respect to $v$ yields
(since $c_1$ does not depend on $v$)
  $$ {\cal H}_v \dot\varphi_v =-Q'(\varphi_v)\eqno(4.5) $$
 and from (4.3) we obtain
$$ \langle {\cal H}_v\dot\varphi_v, \dot\varphi_v\rangle =
  -\ddot d(v) <0.\eqno(4.6)$$
 Putting $y=a_1\chi+p_1$, $p_1\in P$, where $\chi$ is an eigenfunction of
 ${\cal H}_v$ corresponding to
 the negative eigenvalue $-\lambda_0^2$ and  $P$ is the positive subspace of
  ${\cal H}_v$, we obtain
$$\langle{\cal H}_v y, y\rangle =-a_1^2\lambda_0^2+
\langle {\cal H}_vp_1, p_1 \rangle.$$
 Write $\dot\varphi_v=a_0\chi+b_0\varphi_v'+p_0$, $p_0\in P$. From (4.6)
 we have
 $$ 0> \langle {\cal H}_v \dot\varphi_v, \dot\varphi_v\rangle
   = -a_0^2\lambda_0^2+\langle {\cal H}_vp_0, p_0 \rangle $$
 and
$$ \begin{array}{rl}
0 =&{\displaystyle-\langle Q'(\varphi_v), y \rangle =
\langle H_v\dot\varphi_v, y \rangle} \\
=&\langle -a_0\lambda_0^2\chi+H_vp_0, a_1\chi+p_1 \rangle =
-a_0a_1\lambda_0^2 +\langle {\cal H}_vp_0, p_1 \rangle.
  \end{array} $$
 Hence
 $$ \begin{array}{ll}
 \langle {\cal H}_v y, y\rangle & ={\displaystyle  -a_1^2 \lambda_0^2
 +\langle {\cal H}_v p_1, p_1 \rangle \geq -a_1^2 \lambda_0^2+
 \frac{\langle {\cal H}_v p_0, p_1 \rangle^2} {\langle {\cal H}_v p_0,
 p_0\rangle } } \\
 &{\displaystyle = -a_1^2\lambda_0^2+ \frac{a_0^2a_1^2\lambda_0^4}
 {\lambda_0^2 a_0^2-\ddot d(v)} =a_1^2K_1}.
 \end{array} $$
 The rest of the proof is the same as in \cite{5}, pages 310--311.

\vspace{2ex}
\noindent
{\bf Proposition 5.} {\it If Assumption 1 holds, then for any positive
$\varepsilon$ there exists $K>0$ such that for any solution of $(1.1)$
satisfying $u\in{\cal U}_\varepsilon$, $Q(u)=Q(\varphi_v)$, one holds}
$$E(u)-E(\varphi_v)+c_1V(u)-c_1V(\varphi_v)
\geq K||u(.+\alpha(u))-\varphi_v||_1^2.\eqno(4.7)$$

\vspace{1ex}
\noindent
{\bf Proof.} Denote
$\psi=u(\cdot+\alpha(u))-\varphi_v=\mu (\varphi_v-\varphi_v'')+y$
where $\langle\varphi_v-\varphi_v'',y\rangle=0$. By the translation invariant
property of $Q$, we have
  $$ \begin{array}{ll}
    Q(\varphi_v)= Q(u)\!\!  &=Q(\varphi_v)+\langle Q'(\varphi_v), \psi
    \rangle + O(||\psi ||_{1}^2) \\
 \!\! &=Q(\varphi_v)+\langle \varphi_v-\varphi_v'',
    \mu (\varphi_v-\varphi_v'')+y \rangle +O(||\psi ||_{1}^2)\\
\!\! &=Q(\varphi_v)+\mu
    ||\varphi_v-\varphi_v''||_{L^2}^2+O(||\psi ||_{1}^2).
    \end{array} $$
   This implies $\mu=O(||\psi ||_{1}^2)$. Since
   $Q(u)=Q(\varphi_v)$, the identity
   $M''(\varphi_v)={\cal H}_v$ and the Taylor expansion give
     $$
       E(u)-E(\varphi_v)+c_1\int_{0}^{T}{\psi}dx=\frac12\langle {\cal H}_v y, y\rangle
       +o(||\psi ||_{1}^2). $$ From Lemma 1 we have
       $$0=\langle u(\cdot+\alpha(u)), \varphi_v'\rangle = \langle
       \mu (\varphi_v-\varphi_v'')+y+\varphi_v ,
       \varphi_v'\rangle =\langle y, \varphi_v'\rangle.$$
     From the above equality and Lemma 2 we obtain
       $$ E(u)-E(\varphi_v)+c_1\int_{0}^{T}{\psi}dx \geq K||y||_{1}^2+o(||\psi||_{1}^2).
       $$
       This estimate together with
         $$
         ||y||_{1}=||\psi-\mu(\varphi-\varphi'')||_{1}\geq
         ||\psi||_{1}-|\mu|||\varphi-\varphi''||_{1} $$
       yield
         $$E(u)-E(\varphi_v)+c_1\int_{0}^{T}{\psi}dx \geq K||\psi||_{1}^2 $$
       for $||\psi||_1$ sufficiently small. This completes the proof.

\vspace{2ex}
 \noindent
 {\bf Theorem 2.} {\it If Assumption 1 holds, then the travelling-wave
solution  $\varphi_v$ is stable.}

\vspace{1ex}
 \noindent
  {\bf Proof.} Suppose that $\varphi_v$ is unstable. Then there
  exists a sequence of initial data $u_n(0)\in H^1$ and $\eta >0$
  such that
    $$ ||u_n(0)-\varphi_v||_{1} \rightarrow 0 $$
   but
    $$ \sup_{t\in [0,t_0]} \inf_{r \in \Rs} ||u_n(\cdot,
    t)-\varphi_v(\cdot-r)||_{1} \geq \eta ,\eqno(4.8)$$
  where $u_n \in C([0, t_0) ; H^1)$ is a solution of (1.1) with initial data
  $u_n(0)$. Let $t_n\in [0, t_0)$ be the first time so that
 $$ \inf_{r \in \Rs} ||u_n(\cdot, t_n)-\varphi_v(\cdot - r)||_{1}=\eta.$$
  We have
    $$ \begin{array}{ll}
         E(u_n(\cdot, t_n))=E(u_n(0)) \rightarrow E(\varphi_v),\\
         Q(u_n(\cdot, t_n))=Q(u_n(0)) \rightarrow Q(\varphi_v),\\
         V(u_n(\cdot, t_n))=V(u_n(0)) \rightarrow V(\varphi_v).
       \end{array} $$
  Choose a sequence $\psi_n\in H^1$ such that
  $Q(\psi_n)=Q(\varphi_v)$ and
  $||\psi_n-u_n(\cdot, t_n)||_{1} \rightarrow 0$.
 By continuity of $E$ and $V$, $E(\psi_n)\rightarrow E(\varphi_v)$ and
  $V(\psi_n)\rightarrow V(\varphi_v)$. From (4.7) we have
   $$ E(\psi_n)-E(\varphi_v)+c_1(V(\psi_n)-V(\varphi_v)) \geq
   K||\psi_n(\cdot+\alpha(\psi_n))-\varphi_v||_{1}^2.$$
 Therefore $||\psi_n-\varphi_v(\cdot-\alpha(\psi_n))||_{1}
 \rightarrow 0$, which implies
   $$||u_n-\varphi_v(\cdot-\alpha(\psi_n))||_{1}
 \rightarrow 0.$$
 This however contradicts (4.8). The proof of Theorem 2 is complete.

\vspace{2ex}
\noindent
In order to apply Theorem 2 we have, according to Assumption 1, to determine
the sign of the derivative $\ddot d(v)$. Using the same technique as in the
proof of Proposition 2, we obtain the following expression of $\ddot d(v)$
through line integrals.  When $a(u)$ and $b(u)$ are polynomials (or even
rational functions), these are complete Abelian integrals. A lot of methods
have been developed to investigate Abelian integrals, which could be applied
here to study the sign of $\ddot d(v)$.

\vspace{2ex}
\noindent
{\bf Proposition 6.} {\it Assume that the period $T=T(\varphi_0)=\bar T(h)$
is not critical where $h$ is given by $(3.11)$. Then}
$$\ddot d(v)=\frac{W(h)}{4\bar T'(h)},\eqno(4.9)$$ with
$$W(h)=\left(\frac{d}{dh}\oint_{H=h}\frac{dx}{y}\right)\!\!\left(\frac{d}{dh}
\oint_{H=h}\!\!\frac{(x^4+2x^2y^2-\frac13y^4)dx}{y}\right)
-\left(\frac{d}{dh}\oint_{H=h}\!\!\frac{(x^2+y^2)dx}{y}\right)^2.$$

\vspace{2ex}
\noindent
{\bf Proof.} We apply again (3.13) and use the Gelfand-Leray form (in both
directions) to calculate the needed derivatives. As
$$\dot d(v)=\frac12\oint_{H=h}\frac{(x^2+y^2)dx}{y}=
\frac12\oint_{H=h}ydx +\frac12\frac{d}{dh}\oint_{H=h} x^2(v-b(x))ydx$$
one obtains
$$\begin{array}{l}\displaystyle
\ddot d(v)=\frac{\dot h}{2}\frac{d}{dh}\oint_{H=h}\frac{(x^2+y^2)dx}{y}+
\frac12\frac{d}{dv}\left(\oint_{H=h}ydx +\frac{d}{dh}\oint_{H=h}
x^2(v-b(x))ydx\right)\\[4mm]
\displaystyle =\frac{\dot h}{2}\frac{d}{dh}\oint_{H=h}\frac{(x^2+y^2)dx}{y}+
\oint_{H=h}\frac{(x^2-y^2)dx}{4(v-b(x))y}+ \frac{d}{dh}\oint_{H=h}
\left(\frac{x^2y}{2}+\frac{x^2(x^2-y^2)}{4y}\right)dx\\[4mm]
\displaystyle =\frac{\dot h}{2}\oint_{H=h}\frac{(x^2+y^2)dx}{y}
+\frac14\frac{d}{dh}\oint_{H=h} \frac{(x^4+2x^2y^2-\frac13y^4)dx}{y}.
\end{array}$$
Replacing the value of $\dot h$ from (3.16), we come to the needed formula.
$\Box$

\vspace{2ex}
\noindent
{\large\bf 5. Examples.}

\vspace{1ex}
\noindent
{\bf Example 5.1: the BBM equation.} Consider the BBM equation
$$u_t+2\omega u_x+3uu_x-u_{xxt}=0,\quad \omega\in\R\eqno\mbox{\rm (BBM)}$$
which is a particular case of (1.1) with $a(u)=2\omega u+\frac32u^2$
and $b=0$. To apply Theorem 2 to BBM, we have to verify Assumption 1.
Namely, to establish that the corresponding operator ${\cal H}_v$ has
the needed spectral properties (i) and to prove the convexity of $d(v)$.

Let us first mention that (3.5) reduces now to $D \equiv (v-2\omega)^2+6c_1>0$
and:
$$X_0=\frac{v-2\omega+D^{1/2}}{3}, \quad
\Delta=\left(\frac{v-2\omega-D^{1/2}}{3}, \frac{v-2\omega+2D^{1/2}}{3}\right)
\quad \mbox{\rm if}\quad v>0,$$
$$X_0=\frac{v-2\omega-D^{1/2}}{3}, \quad
\Delta=\left(\frac{v-2\omega-2D^{1/2}}{3}, \frac{v-2\omega+D^{1/2}}{3}\right)
\quad \mbox{\rm if}\quad v<0.$$
By the definition of $c_1, c_2$ and $U(s)$ one obtains in the considered case
$$\begin{array}{rl}
U(s)&\equiv \displaystyle \frac{1}{v}(\varphi_0-s)[s^2+(\varphi_0+2\omega-v)s
-(2\varphi_0^2+(2\omega-v)\varphi_0+2v\varphi_2)]\\[4mm]
&\displaystyle=\frac{(s-\varphi_0)(\varphi_1-s)(s+\varphi_1+\varphi_0+2\omega-v)}{v}.
\end{array}$$
We note that the last equality is a consequence of Proposition 1 (ii) which
implies that $U(\varphi_1)=U(\varphi_0)=0$. To obtain an explicit formula
for the travelling wave $\varphi_v$, we substitute
$\sigma=\varphi_0+(\varphi_1-\varphi_0)z^2$, $z>0$ in order to express the
integral in (3.6) as an elliptic integral of the first kind in a Legendre
form. If $v<0$, one obtains
$$\int_0^{Z(s)}\frac{dz}{\sqrt{(1-z^2)(1-k^2z^2)}}=\alpha(s-\xi)$$
where
$$Z(s)=\sqrt{\frac{\varphi_v(s)-\varphi_0}{\varphi_1-\varphi_0}},\quad
k^2=-\frac{\varphi_1-\varphi_0}{\varphi_1+2\varphi_0+2\omega-v}, \quad
\alpha=\sqrt{\frac{\varphi_1+2\varphi_0+2\omega-v}{4v}}.$$
Thus we get the expression
$$\varphi_v(s)=\varphi_0+(\varphi_1-\varphi_0)sn^2(\alpha(s-\xi);k).\eqno(5.1)$$
Similarly, in the case $v>0$ we obtain (with the same $Z$)
$$\int_0^{Z(s)}\frac{dz}{\sqrt{(1-z^2)(k'^2+k^2z^2)}}=\alpha(s-\xi)$$
where
$$k^2=\frac{\varphi_1-\varphi_0}{\varphi_0+2\varphi_1+2\omega-v}, \quad
k^2+k'^2=1,\quad
\alpha=\sqrt{\frac{\varphi_0+2\varphi_1+2\omega-v}{4v}},$$
and the expression for $\varphi_v$
$$\varphi_v(s)=\varphi_0+(\varphi_1-\varphi_0)cn^2(\alpha(s-\xi);k).\eqno(5.2)$$

To calculate the period of $\varphi_v$, we use (3.7) and the same procedure
as above. In this way we get in both cases
$$T= 2\int_{\varphi_0}^{\varphi_1}\frac{d\sigma}{\sqrt{U(\sigma)}}=
 \frac{2}{\alpha}\int_0^1\frac{dz}{\sqrt{(1-z^2)(1-k^2z^2)}}=
 \frac{2K(k)}{\alpha}.$$

We return to the operator ${\cal H}_v$ defined by (4.4) which now has the form
$${\cal H}_v=-v\partial_x^2+v-2\omega-3\varphi_v \eqno(5.3)$$
where $\varphi_v$ is determined by (5.1) or (5.2). Take $v>0$ and consider the
spectral problem
$$\begin{array}{l}
{\cal H}_v\psi=\lambda\psi,\\
\psi(0)=\psi(T),\;\psi'(0)=\psi'(T).\end{array}\eqno(5.4)$$ We
will denote the operator just defined again by ${\cal H}_v$. It is
a self-adjoint operator acting in $H^2([0,T])$. From the Floquet
theory applied to (5.4) it follows \cite{1} that its spectrum is
purely discrete,
$$\lambda_0<\lambda_1\leq\lambda_2<\lambda_3\leq\lambda_4<\ldots \eqno(5.5)$$
where $\lambda_0$ is always a simple eigenvalue. If $\psi_n(x)$ is the
eigenfunction corresponding to $\lambda_n$, then
$$\begin{array}{l}
\psi_0 \;\;\mbox{\rm has no zeroes in}\;\; [0,T];\\
\psi_{2n+1},\;\psi_{2n+2} \;\;\mbox{\rm have each just} \;\; 2n+2
\;\;\mbox{\rm zeroes in}\;\; [0,T).
\end{array}\eqno(5.6)$$

\vspace{2ex}
\noindent
{\bf Proposition 7.} {\it The linear operator ${\cal H}_v$ defined by
$(5.3)-(5.4)$ has the following spectral properties for $v>0$:}

\vspace{1ex}
\noindent
(i) {\it The first three eigenvalues of ${\cal H}_v$ are simple.}

\vspace{1ex}
\noindent
(ii) {\it The second eigenvalue of ${\cal H}_v$ is $\lambda_1=0$.}

\vspace{2ex} \noindent {\bf Proof.} By (3.1), ${\cal
H}_v\varphi'_v=0$, hence  $\psi=\varphi_v'$ is an eigenfunction
corresponding to zero eigenvalue. By Proposition 1 (iii)
$\varphi'$ has just two zeroes in $[0,T)$ and therefore by (5.6)
either $0=\lambda_1<\lambda_2$ or $\lambda_1<\lambda_2=0$ or
$\lambda_1=\lambda_2=0$. We are going to verify that only the
first possibility $0=\lambda_1<\lambda_2$ can occur. From the
definition of $k$ and $\alpha$ one obtains that
$$\varphi_0+2\varphi_1+2\omega-v=4v\alpha^2,\quad
\varphi_1-\varphi_0=4vk^2\alpha^2.$$ Then using (5.2) we get
$$\begin{array}{rl}
{\cal H}_v & =-v\partial_x^2+v-2\omega
-3\varphi_0-3(\varphi_1-\varphi_0)cn^2(\alpha x;k)\\[2mm]
 & =-v\partial_x^2+v-2\omega
-3\varphi_1+3(\varphi_1-\varphi_0)sn^2(\alpha x;k)\\[2mm]
& = -v\partial_x^2-v\alpha^2[4k^2+4-12k^2sn^2(\alpha x;k)]\\[2mm]
& = v\alpha^2[-\partial_y^2 -4k^2-4+12k^2sn^2(y;k)]\equiv
 v\alpha^2\Lambda \end{array}$$
where $y=\alpha x$. The operator $\Lambda$ is related to
Hill's equation with Lam\'e potential
$$\Lambda w=-\frac{d^2}{dy^2}w+[12k^2sn^2(y;k)-4k^2-4]w=0$$
and its spectral properties in the interval $[0,2K(k)]$ are well
known \cite{An,AnSa,In}. The first
three (simple) eigenvalues and corresponding periodic
eigenfunctions of $\Lambda$ are
$$\begin{array}{l}
\mu_0=k^2-2-2\sqrt{1-k^2+4k^4}<0, \\
\psi_0(y)=dn(y;k)[1-(1+2k^2-\sqrt{1-k^2+4k^4})sn^2(y;k)]>0,\\[2mm]
\mu_1=0,\\
\psi_1(y)=dn(y;k)sn(y;k)cn(y;k)=\frac12(d/dy)sn^2(y;k),\\[2mm]
\mu_2=k^2-2+2\sqrt{1-k^2+4k^4}>0,\\
\psi_2(y)=dn(y;k)[1-(1+2k^2+\sqrt{1-k^2+4k^4})sn^2(y;k)].\end{array}$$
As the eigenvalues of ${\cal H}_v$ and $\Lambda$ are related by
$\lambda_n=v\alpha^2\mu_n$ we conclude that for $v>0$, the first
three eigenvalues of (5.3)-(5.4) are simple and moreover $\lambda_0<0$,
$\lambda_1=0$, $\lambda_2>0$. The corresponding eigenfunctions are
$\psi_0(\alpha x)$, $\psi_1(\alpha x)=const.\varphi'_v(x)$ and
$\psi_2(\alpha x)$. $\Box$

\vspace{1ex}
What about the sign of $\ddot d(v)$, it is easily seen it is positive
when $\omega=0$ (see the end of this section). For $\omega\neq 0$, the
proof is much more long and complicated. It can be done by following the
procedure we propose in the next subsection. For this reason, we are not
going to investigate the case $\omega\neq 0$ in the present paper.

\vspace{2ex}
\noindent
{\bf Example 5.2: the modified BBM equation.} Our second example is
concerned with the periodic travelling-wave solutions of the equation
$$u_t+2wu_x+\beta (u^3)_x-u_{xxt}=0 \eqno\mbox{\rm (mBBM)} $$
where $\omega, \beta \in \R$ and $\beta\neq 0$. For this case,
$a(u)=2wu+\beta u^3$  and $b(u)=0$ in (1.1). For definiteness, we take
$\varphi_v=\varphi(x-vt)$ where $v>0$. The Hamiltonian flow in (3.4) is
generated by
$$H(X,Y)=\frac{v}{2}Y^2+\frac{\beta}{4}X^4
+\left(\omega-\frac{v}{2}\right)X^2-c_1X.$$
Our plan is to study here the "symmetric" case $c_1=0$. The general case
could then be considered as a perturbation of the symmetric one (at least for
$c_1$ small). It is well known that there are three cases related to the
symmetric Hamiltonian
$$Y^2+\frac{\beta}{2v}X^4+\left(\frac{2\omega}{v}-1\right)X^2, \quad v>0,$$
see e.g. \cite{6}:

\vspace{1ex}
(i) global center:  $\beta>0$, $2\omega>v$;

\vspace{1ex}
(ii) truncated pendulum: $\beta<0$, $2\omega>v$;

\vspace{1ex}
(iii) Duffing oscillator: $\beta>0$, $2\omega<v$,

\vspace{1ex}
\noindent
with three topologically different phase portraits.
There is one continuous family of periodic orbits in cases (i) and (ii)
and three families (left, right, outer) in (iii) (we advice the reader to
draw a picture). This also holds true for all $c_1$ in case (i) and for
$c_1^2<\frac{4}{27}\beta^{-1}(v-2\omega)^3$ in (ii), (iii). In the symmetric
case $c_1=0$ we deal with, the periodic solutions exist if and only if
$\varphi_0\in\Delta^-$ (see Proposition 1) where:

\vspace{1ex}
$\Delta^-=(-\infty,0)$ in case (i);

\vspace{1ex}
$\Delta^-=(-(\frac{v-2\omega}{\beta})^{1/2},0)$ in case (ii);

\vspace{1ex}
$\Delta^-=(-(\frac{2v-4\omega}{\beta})^{1/2},-(\frac{v-2\omega}{\beta})^{1/2})$
in case (iii), left family;

\vspace{1ex}
$\Delta^-=(0,(\frac{v-2\omega}{\beta})^{1/2})$ in case (iii), right family;

\vspace{1ex}
$\Delta^-=(-\infty, -(\frac{2v-4\omega}{\beta})^{1/2})$ in case (iii), outer family.

\vspace{1ex}
The respective functions $U(s)$ take the form:
$$\begin{array}{ll}
U(s)=\frac{\beta}{2v}(\varphi_0^2-s^2)(\frac{4\omega-2v}{\beta}+\varphi_0^2+s^2)
 & \mbox{\rm in case (i) and case (iii), outer family,}\\
U(s)=-\frac{\beta}{2v}(\varphi_0^2-s^2)(\frac{2v-4\omega}{\beta}-\varphi_0^2-s^2)
 & \mbox{\rm in case (ii),}\\
U(s)=\frac{\beta}{2v}(s^2-\varphi_1^2)(\varphi_0^2-s^2)
 & \mbox{\rm in case (iii), left family,}\\
U(s)=\frac{\beta}{2v}(s^2-\varphi_0^2)(\varphi_1^2-s^2)
 & \mbox{\rm in case (iii), right family}.\end{array}$$
In a similar way as we have done in the (BBM) case, we can
use formula (3.6) to calculate explicitly $\varphi_v$. By an appropriate
change of the variables in (3.6), one can express the solution through
standard elliptic integrals (we omit the details). Up to a translation of
the argument, we have:
$$\begin{array}{llll}
\varphi_v(s)=\varphi_0 cn(\alpha s;k),
& \alpha=\sqrt\frac{2\omega-v+\beta\varphi_0^2}{v},
& k=\sqrt\frac{\beta\varphi_0^2}{4\omega-2v+2\beta\varphi_0^2},&\mbox{\rm (i) and (iii) outer,}\\[3mm]
\varphi_v(s)=\varphi_0 sn(\alpha s;k),
& \alpha=\sqrt\frac{4\omega-2v+\beta\varphi_0^2}{2v},
&  k=\sqrt\frac{\beta\varphi_0^2}{2v-4\omega-\beta\varphi_0^2},&\mbox{\rm (ii),}\\[3mm]
\varphi_v(s)=\varphi_0  dn(\alpha s;k),
& \alpha=\sqrt\frac{\beta\varphi_0^2}{2v},
&  k=\sqrt\frac{4\omega-2v+2\beta\varphi_0^2}{\beta\varphi_0^2},&\mbox{\rm (iii) left,}\\[3mm]
\varphi_v(s)=\varphi_1  dn(\alpha s;k),
& \alpha=\sqrt\frac{\beta\varphi_1^2}{2v},
&  k=\sqrt\frac{4\omega-2v+2\beta\varphi_1^2}{\beta\varphi_1^2},&\mbox{\rm (iii) right.}
\end{array}$$
These formulas of $\alpha$ and $k$ yield the following expressions and range $I$
for the period:
$$\begin{array}{llll}
T=4\sqrt\frac{v}{2\omega-v}\sqrt{1-2k^2}K(k),& k\in(0,\frac{1}{\sqrt{2}}),
& I=(0,2\pi\sqrt\frac{v}{2\omega-v}), &\mbox{\rm (i),}\\[3mm]
T=4\sqrt\frac{v}{2\omega-v}\sqrt{1+k^2}K(k),& k\in(0,1),
& I=(2\pi\sqrt\frac{v}{2\omega-v},\infty), &\mbox{\rm (ii),}\\[3mm]
T=2\sqrt\frac{v}{v-2\omega}\sqrt{2-k^2}K(k),& k\in(0,1),
& I=(2\pi\sqrt\frac{v}{2v-4\omega},\infty), &\mbox{\rm (iii) left and right,}\\[3mm]
T=4\sqrt\frac{v}{v-2\omega}\sqrt{2k^2-1}K(k),&
k\in(\frac{1}{\sqrt2},1), & I=(0,\infty), &\mbox{\rm (iii)
outer.}\end{array}$$
Indeed, the formulas just derived imply that
$T=T(k)$ is strictly decreasing in case (i) and strictly
increasing in the other cases. In fact, for (i), (ii) and (iii)
outer this follows already from (3.11) and \cite{4}. As far as
cases (iii) left and right are concerned, now the result follows from
$$\frac{d}{dk}(\sqrt{2-k^2}K(k))=\frac{(2-k^2)K'-kK}{\sqrt{2-k^2}}=
\frac{K'+E'}{\sqrt{2-k^2}}>0.$$ On the other hand, in all cases
one has $dk/d\varphi_0\neq 0$. Therefore, given $T\in I$, the
condition holds in order to determine $\varphi_0$ by the implicit
function theorem so that the respective $\varphi$ would have a
period $T$.

Finally, by using the above formulas one easily obtains that
$${\cal H}_v=-v\partial_x^2+v-2\omega-3\beta\varphi_v^2
=v\alpha^2[-\partial_y^2+6k^2sn^2(y; k)+m]$$
where $y=\alpha x$ and
$m=-1-4k^2$ in cases (i) and (iii) outer,
$m=-4-k^2$ in cases (iii) left and right,
$m=-1-k^2$ in case (ii), respectively.

\vspace{2ex}
\noindent
{\bf Lemma 3.} {\it The first five eigenvalues of the operator
$\Lambda$ defined by the differential expression
$\Lambda= -\partial_y^2+6k^2sn^2(y;k)$, with
periodic boundary conditions on $[0, 4K(k)]$, are simple. These
eigenvalues and their respective eigenfunctions are:}
$$\begin{array}{ll}
\mu_0=2+2k^2-2\sqrt{1-k^2+k^4}, &\psi_0(y)=1-(1+k^2-\sqrt{1-k^2+k^4})sn^2(y;k),\\[1mm]
\mu_1=1+k^2, &\psi_1(y)=cn(y;k)dn(y;k)=sn'(y;k),\\[1mm]
\mu_2=1+4k^2, &\psi_2(y)=sn(y;k)dn(y;k)=-cn'(y;k),\\[1mm]
\mu_3=4+k^2, &\psi_3(y)=sn(y;k)cn(y;k)=-k^{-2}dn'(y;k),\\[1mm]
\mu_4=2+2k^2+2\sqrt{1-k^2+k^4}, &\psi_4(y)=1-(1+k^2+\sqrt{1-k^2+k^4})sn^2(y;k).
\end{array}$$
{\bf Proof.} The equalities $\Lambda \psi_n(y)=\mu_n\psi_n(y)$, $0\leq n\leq 4$
are established by calculation. By (5.5), (5.6) and the properties of elliptic
functions, $\mu_n$ are simple and the rest of the spectrum lies in the interval
$(\mu_4,\infty)$. $\Box$

\vspace{2ex}
\noindent
{\bf Corollary 1.} {\it The first three eigenvalues of the operator $\Lambda$,
equipped with periodic boundary conditions on $[0, 2K(k)]$, are simple and
equal to $\mu_0$, $\mu_3$, $\mu_4$.}

\vspace{2ex}
\noindent
By Lemma 3 and its corollary, $\varphi_v'$ is the second eigenfunction
of the operator ${\cal H}_v$ in the cases: truncated pendulum and
Duffing oscillator (left and right). For these cases, it makes sense
to investigate the sign of $\ddot d(v)$ which we do in the next proposition.

\vspace{2ex}
\noindent
{\bf Proposition 8.} {\it Assume that $T\in I=(T_0,\infty)$ and
$\varphi_v=\varphi(x-vt)$, $v>0$, is the periodic travelling-wave solution
having a minimal period $T$. Then:}

\vspace{1ex}
\noindent
(i) {\it In the truncated pendulum case one has $\ddot d(v)<0$.}

\vspace{1ex}
\noindent
(ii) {\it In the left $($right$)$ Duffing oscillator case, if
$3v^2-8\omega^2\geq 0$,
then $\ddot d(v)>0$. If $3v^2-8\omega^2<0$ and $2v^2-2\omega v-\omega^2>0$,
then there is $T_{max}\in (T_0,\infty)$ depending only on the ratio $\omega/v$, such that
$\ddot d(v)>0$ for $T>T_{max}$ and $\ddot d(v)<0$ for $T\in (T_0,T_{max})$.
If $2v^2-2\omega v-\omega^2\leq 0$, then $\ddot d(v)<0$.}

\vspace{2ex}
\noindent
{\bf Proof.} We are going to use formula (4.9). In the example we deal with,
one has
$$y^2=U(x,h)=\frac{2h}{v}-\frac{2\omega-v}{v}x^2-\frac{\beta}{2v}x^4,
\quad
h=\frac{2\omega-v}{2}\varphi_0^2+\frac{\beta}{4}\varphi_0^4.$$
Given a nonnegative even integer $n$, denote
$I_n(h)=\oint_{H=h}x^nydx$. It is well known that the linear space
of integrals $\{I_n(h), n$ even$\}$ forms a polynomial $R[h]$
module with two generators, $I_0(h)$ and $I_2(h)$. Moreover, $I_0$
and $I_2$ satisfy a Picard-Fuchs system of dimension two. This
implies that the ratio $R(h)=I_2'(h)/I_0'(h)$ satisfies a Riccati
equation. We shall use these facts to express $\ddot d(v)$ as a
quadratic form with respect to $I_0'$, $I_2'$ with polynomial
coefficients in $h$ and use the properties of the Riccati equation
to determine  the sign of $\ddot d(v)$.The procedure might seem
too long, but it is universal (at least in the case when $a(u)$
and $b(u)$ are polynomials) and therefore applicable to many other
cases.

Let us first express $\ddot d(v)$ through the integrals $I_n$.
Below, we will denote for short the derivatives with respect to
$h$  by $I_n'$, $I_n''$ etc.  Using the first equality in (3.13)
with $b=0$, we obtain
$$I'_n(h)=\oint_{H=h}\frac{x^ndx}{vy}.$$
On the other hand,
$$\oint_{H=h}y^3dx=\oint_{H=h}U(x,h)ydx=
\frac{2h}{v}I_0-\frac{2\omega-v}{v}I_2-\frac{\beta}{2v}I_4.$$
By  using these expressions, we obtain that
$$\ddot d(v)=\frac{v}{4I_0''}\left [I_0''\left(I_4'+\frac{\beta}{6v^2}I_4
+\frac{2\omega+5v}{3v^2}I_2-\frac{2h}{3v^2}I_0\right)'-\left(\frac{1}{v}I_0'
+I_2''\right)^2\right].\eqno(5.7)$$
For reader's convenience, below we proceed to derive the relations between
integrals $I_n$ and the Picard-Fuchs system satisfied by $I_0$ and $I_2$.

\vspace{2ex}
\noindent
{\bf Lemma 4.} (i) {\it The following relations hold:}
$$(n+6)\beta I_{n+3}+(2n+6)(2\omega-v)I_{n+1}=4nhI_{n-1},\quad n=1,3,5,
\ldots.\eqno(5.8)$$
(ii) {\it
The integrals $I_0$ and $I_2$ satisfy the system}
$$\begin{array}{l}
4hI_0'-(2\omega-v)I_2'=3I_0,\\[2mm]
\displaystyle -\frac{4(2\omega-v)}{3\beta}hI'_0+\left(4h+\frac{4(2\omega-v)^2}
{3\beta}\right)I_2'=5I_2.\end{array}\eqno(5.9)$$

\vspace{1ex}
\noindent
{\bf Proof.} (i). Integrating by parts, we obtain the identity
$$\oint_{H=h}[x^nU'(x)+{\textstyle\frac23}nx^{n-1}U(x)]ydx=0.\eqno(5.10)$$
Indeed,
$$\begin{array}{l}\displaystyle
\oint_{H=h}x^nU'(x)ydx=\oint_{H=h}x^nydy^2=
{\textstyle\frac23}\oint_{H=h}x^ndy^3=-{\textstyle\frac23}n\oint_{H=h}x^{n-1}y^3dx\\[4mm]
\displaystyle=-{\textstyle\frac23}n\oint_{H=h}x^{n-1}U(x)ydx.\end{array}$$
Identity (5.10) is equivalent to (5.8).

(ii) Similarly, one has
$$\begin{array}{l}\displaystyle
\oint_{H=h}\frac{x^nU'(x)dx}{y}=\oint_{H=h}\frac{x^ndy^2}{y}=
\oint_{H=h}2x^ndy=-2n\oint_{H=h}x^{n-1}ydx,\\[4mm]
\displaystyle\oint_{H=h}\frac{x^nU(x)dx}{y}=\oint_{H=h}x^nydx.
\end{array}\eqno(5.11)$$
Rewriting these identities by means of $I_n$, we come to the formulas
$$\begin{array}{l}
\beta I'_{n+3}+(2\omega-v)I'_{n+1}=nI_{n-1},\\
-\beta I'_{n+4}-2(2\omega-v)I'_{n+2}+4hI'_n=2I_n.\end{array}\eqno(5.12)$$
The last two relations imply
$$4hI'_n-(2\omega-v)I'_{n+2}=(n+3)I_n.\eqno(5.13)$$
Using (5.13) with $n=0,2$, we obtain the system
$$\begin{array}{l}
4hI'_0-(2\omega-v)I'_2=3I_0,\\
4hI'_2-(2\omega-v)I'_4=5I_2.\end{array}$$
We remove $I_4'$ from the last equation by using the first equation in (5.12).
As a result one obtains (5.9). $\Box$

After proving Lemma 4, we return to our main goal.
Using the identities
$$I_4=\frac{4h}{7\beta}I_0-\frac{8(2\omega-v)}{7\beta}I_2,\quad
I_4'=\frac{1}{\beta}I_0-\frac{2\omega-v}{\beta}I_2'.$$
and the first equation in (5.9), we remove $I_4$, $I'_4$ and $I_0$
from (5.7). The result is
$$\ddot d(v)=\frac{v}{4I_0''}\left [I_0''\left(
\frac{3v^2-4\beta h}{3\beta v^2}I_0'+\frac{2\omega+5v}{3v^2}I_2'
-\frac{2\omega-v}{\beta}I_2''\right)-\left(\frac{1}{v}I_0'
+I_2''\right)^2\right].\eqno(5.14)$$
Now, we differentiate (5.9) and determine the second derivatives
from the obtained system:
$$\begin{array}{l}
{\cal D}(h)I_0''=-4hI_0'+(2\omega-v)I_2',\\[2mm]
\displaystyle {\cal D}(h)I_2''=\frac{4(2\omega-v)}{\beta}hI_0'+4hI_2',\\[2mm]
\displaystyle {\cal D}(h)=16h^2+\frac{4(2\omega-v)^2}{\beta}h.
\end{array}\eqno(5.15)$$
Replacing in (5.14) and performing some direct calculations to simplify the
result, we derive the formula
$$\ddot d(v)=\frac{1}{9vI_0}\left
[\left(8h^2+\frac{12\omega^2}{\beta}h\right)I_0'^2+(4\omega+10v)hI_0'I_2'
+(2v^2-2\omega v-\omega^2)I_2'^2\right].$$
By (5.15), the ratio $R(h)=I_2'(h)/I_0'(h)$ satisfies the Riccati equation
$${\cal D}(h)R'=\frac{4(2\omega-v)}{\beta}h+8hR-(2\omega-v)R^2.\eqno(5.16)$$
In order to remove the parameters from (5.16), we take new variables
$\bar h$, $\bar R$ through
$$h=-\frac{(2\omega-v)^2}{8\beta}(\bar h+1),\quad R(h)=-\frac{2\omega-v}{2\beta}
(\bar R(\bar h)+1).$$
Thus we obtain the final formulas
$$\ddot d(v)=\frac{(2\omega-v)^2I_0'^2(h)}{72v\beta^2I_0(h)}w(\bar h, \bar R)$$
where (below we will omit thoroughly the bars)
$$w(h,R)=(4v^2-4\omega v-2\omega^2)R^2+(4\omega^2+8\omega v-5v^2)hR
+(2\omega-v)^2h^2 +3v^2(R-h)-6\omega^2\eqno(5.17)$$
and $R=R(h)$ satisfies the equation (and related system with respect to a
dummy variable)
$$4(1-h^2)R'=1-2hR+R^2,\qquad\qquad \begin{array}{l}\dot h=4(1-h^2),\\
\dot R=1-2hR+R^2.\end{array}\eqno(5.18)$$ Since $v>0$ and $I_0>0$
(this is the area integral), the sign of $\ddot d(v)$ is
determined by $w$. The conic curve $w(h,R)=0$ divides the
$\R^2$-plane into two parts $W_+$ and $W_-$ according to the sign
of $w(h,R)$. We have to identify the curve $\Gamma$ in the phase
portrait of (5.18) which corresponds to $I_2'/I_0'$ and determine
its location with respect to $W_+$ and $W_-$. In the case when
$\ddot d(v)$ changes sign, the main problem will be to prove that
the conic curve intersects $\Gamma$ at most once.

It is easy to see that system (5.18) has two critical points $(1,1)$ and
$(-1,-1)$ which are saddle-nodes. They are connected by two separatrix
trajectories, upper $\Gamma_u$ and lower $\Gamma_l$,
please draw a picture.
Obviously, the phase portrait of (5.18) is symmetric with respect to the
origin.

\vspace{2ex}
\noindent
{\bf Lemma 5.} {\it In $(5.18)$, the trajectory corresponding to the truncated
pendulum case is $\Gamma_l$. The trajectory corresponding to the left and right
Duffing oscillator cases is $\Gamma_u$.}

\vspace{1ex}
\noindent
{\bf Proof.} If $\varphi_0\in\Delta_-$, then  $\varphi_0^2\in
(0,\frac{v-2\omega}{\beta})$ in the truncated pendulum and right Duffing
oscillator cases and
$\varphi_0^2\in(\frac{v-2\omega}{\beta},\frac{2v-4\omega}{\beta})$
in the left case. Correspondingly, one obtains
$$\begin{array}{l}\displaystyle
h\in \left(0,-\frac{(2\omega-v)^2}{4\beta}\right)=(h_c,h_s)\;\;
\mbox{\rm in the truncated pendulum case,}\\[3mm]
\displaystyle h\in \left(-\frac{(2\omega-v)^2}{4\beta},0\right)=(h_c,h_s)\;\;
\mbox{\rm in the left and right Duffing oscillator cases,}\end{array}$$
where $h_c$ is the Hamiltonian level corresponding to a center
and $h_s$ -- to a saddle.
Moreover, $R(h)=I_2'(h)/I_0'(h)$ is analytic in a neighborhood of $h=h_c$
and, by the mean value theorem,
$$
\lim\limits_{h\to h_c}R(h)=\lim\limits_{h\to h_c}\frac{I_2(h)}{I_0(h)}=X_0^2.
$$
We recall that $X_0$ is the abscissa of the center and $X_0=0$ in the truncated
pendulum case, $X_0^2=\frac{v-2\omega}{\beta}$ in the left and right Duffing
oscillator case. On its turn, by Picard-Lefschetz theory (\cite{8}, Chapter 3), near
$h=h_s$ the integral $I_n(h)$ has the expansion
$$I_n(h)=I_n(h_s)+\alpha_n(h-h_s)\log|h-h_s| +\beta_n(h-h_s)
+\gamma_n (h-h_s)^2\log|h-h_s|+\ldots, $$
where $\alpha_0\neq 0$. Therefore $R(h)=I_2'(h)/I_0'(h)$ is bounded
near $h=h_s$. In terms of the new coordinates $(\bar h, \bar R)$, all this
means that the phase trajectory of (5.18) corresponding to the truncated
pendulum case is $(h,R(h))$ where $-1<h<1$, $R(h)\to\pm1$ as $h\to\pm1$,
and $R(h)$ is analytic near $h=-1$. The unique phase curve with these
properties is $\Gamma_l$. Similarly, the phase trajectory of (5.18)
corresponding to the left and right Duffing oscillator cases has the same
ends but is analytic near $h=1$, hence it is $\Gamma_u$. $\Box$

\vspace{1ex}
Below, we list some properties of the separatrices $\Gamma_l$, $\Gamma_u$ and
the conic curve $w=0$. Apart of the separatrices, the conic curve depends on
one parameter $\omega/v$ and undergoes several bifurcations when $\omega/v$
varies.

\vspace{2ex}
\noindent
{\bf Lemma 6.} {\it The separatrices $\Gamma_u$ and $\Gamma_l$ have
the following properties:}

\vspace{1ex}
\noindent
(i) {\it $\Gamma_u$ is increasing and concave, $\Gamma_l$ is increasing and
convex.}

\vspace{1ex}
\noindent
(ii) {\it The tangent to $\Gamma_u$ at $(-1,-1)$ is $h=-1$,
the tangent to $\Gamma_l$ at $(1,1)$ is $h=1$.}

\vspace{1ex}
\noindent
(iii) {\it The tangent to $\Gamma_u$ at $(1,1)$ and
the tangent to $\Gamma_l$ at $(-1,-1)$ have a slope $\frac14$.}

\vspace{2ex}
\noindent
{\bf Proof.} The proof easily follows from (5.18). Claim (iii) is a consequence
of the analyticity at the corresponding point. To establish (ii), we use the
expansions containing logarithmic terms above. They imply that the asymptotic
expansion near $(-1,-1)$ of the non-analytic trajectories has the form
$-1+\mu/\log(h+1)+\ldots$. We determine $\mu$ from (5.18) and obtain that on
$\Gamma_u$ one holds
$$R(h)=-1-\frac{8}{\log(h+1)}+O\left(\log^{-2}(h+1)\right).\eqno(5.19)$$
A similar formula holds for $\Gamma_l$, hence (ii) is proved. Finally,
in the strip $|h|<1$ one has $1-2hR+R^2>0$, therefore $R'>0$ and all
trajectories increase. Differentiating with respect to $h$ the Riccati
equation in (5.18), we determine the curve of inflection points $R''(h)=0$
in the phase portrait, which is $(R-h)(R^2+2hR-3)=0$. As $R'=\frac14$
on the line $R=h$, by (ii) and (iii) $\Gamma_u$ and $\Gamma_l$ cannot intersect
this line. Hence, $\Gamma_u$ lies in the domain corresponding to
the concave trajectories, and $\Gamma_l$ -- to the convex ones. $\Box$

The proof of the statements listed below is obvious.

\vspace{2ex}
\noindent
{\bf Lemma 7.} {\it The conic curve $w(h,R)=0$ has the following properties:}

\vspace{1ex}
\noindent
(i) {\it It goes through the critical points $(1,1)$ and $(-1,-1)$
of $(5.18)$ and $w(h,h)=6\omega^2(h^2-1)$.}

\vspace{1ex}
\noindent
(ii) {\it For $\omega=0$ it degenerates into $v^2(R-h)(4R-h+3)=0$.
For $\omega^2+2\omega v-2v^2=0$ it degenerates into
$v^2h[(R-1+(7\mp4\sqrt3)(h-1)]=0$.}

\vspace{1ex}
\noindent
(iii) {\it If $(h, R(h))$ are the local coordinates near $(1,1)$, then
$R'(1)=1-\frac{2\omega^2}{v^2}$,
$R''(1)=\frac{8\omega^4(\omega^2+2\omega v-2v^2)}{3v^6}$. }

\vspace{1ex}
\noindent
(iv) {\it If $(h(R), R)$ are the local coordinates near $(-1,-1)$
and $\omega\neq 0$, then $h'(-1)=0$,
$h''(-1)=\frac{2v^2-2\omega v-\omega^2}{3\omega^2}$.}

\vspace{2ex}
\noindent
After the preparation done above, we proceed to prove statements (i) and (ii)
of Proposition 8. To prove (i), let us denote by $\Omega$ the open
triangle in the $(h,R)$-plane having vertices at $(1,1)$, $(-1,-1)$ and
$(1,-\frac12)$. By Lemma 6, $\Gamma_l\subset \Omega$. On its turn,
$\Omega\subset W_-$. This holds because by Lemma 7 one has $w(h,h)<0$
for $|h|<1$, $w(1,1)=w(-1,-1)=0$, $R'(1)<1$, $h'(-1)=0$, and because
$w(1,-\frac12)=-\frac92\omega(\omega+2v)<0$ (recall that $2\omega>v>0$
in the truncated pendulum case). Proposition 8 (i) is established.

The proof of (ii) is more complicated. In this case, $2\omega<v$.
Assume first that $\omega^2+2\omega v-2v^2 > 0$. Then, by Lemma 7 (iii), (iv)
and convexity, the conic curve is a hyperbola. One of its branches is contained
in the half-plane $h\leq -1$, the other branch lies above its tangent line at
$(1,1)$, namely $R-1=R'(1)(h-1)$. As $R'(1)<0$ in the considered case, by
Lemma 6 (i) and Lemma 7 (i) we conclude that $\Gamma_u\subset W_-$. The same
proof goes in the degenerate case $\omega^2+2\omega v-2v^2=0$.

The case $\omega^2+2\omega v-2v^2<0$ is more delicate.  If
$\omega=0$, then Lemma 6 (i) and Lemma 7 (ii) imply that
$\Gamma_u\subset  W_+$. Below we will take $\omega\neq 0$.
Consider the branch $R=r(h)$ of the conic curve going through both
$(1,1)$ and $(-1,-1)$. Writing $w(h,R)=\delta_0R^2+\delta_1
R+\delta_2$, $\delta_0>0$,  one obtains
$$r(h)=\frac{-\delta_1+\sqrt{D}}{2\delta_0},
\quad D=\delta_1^2-4\delta_0\delta_2>0, \quad |h|<1.\eqno(5.20)$$
Below we will denote this branch by $C$. By (5.19) and Lemma 7 (iv) we have
near $h=-1$
$$R+1\sim-\frac{8}{\log(h+1)}\;\; \mbox{\rm on}\;\; \Gamma_u,\quad
R+1\sim\sqrt\frac{6\omega^2(h+1)}{2v^2-2\omega v-\omega^2}
\;\;\mbox{\rm on}\;\; C.$$ This yields that for $h$ close to $-1$,
$\Gamma_u$ is placed above $C$. Similarly, by Lemma 6 (iii) and
Lemma 7 (iii) one obtains that near $h=1$, $\Gamma_u$ lies above
$C$ if $3v^2-8\omega^2\geq 0$ and below $C$ otherwise. Therefore,
to finish the proof of Proposition 7, we have to establish the
following: 1) $\Gamma_u$ is entirely placed above $C$ if
$3v^2-8\omega^2\geq 0$, and 2) $\Gamma_u$ intersects the conic
just once if $3v^2-8\omega^2<0$. In the first case, we would have
$\Gamma_u\subset W_+$, while in the second one, the part of
$\Gamma_u$ near $h=1$ would be in $W_-$ and the remaining part --
in $W_+$. Unfortunately, both $\Gamma_u$ and $C$ are concave and
it is not so easy to determine the number of their intersections.

Below, we proceed to determine the number of contact points that $C$ has
with the vector field (5.18). Because of the type of critical points, in our
case the number of intersections is less than or equal to the number of
contact points. As well known, the equation of the contact points is given by
$$\frac{d}{ds}(R-r(h))_{|R=r(h)}=[\dot R-\dot h r'(h)]_{|R=r(h)}=
1-2hr(h)+r^2(h)-4(1-h^2)r'(h)=0.\eqno(5.21)$$
We replace $r'=-(\delta_1'r+\delta_2')/(2\delta_0r+\delta_1)$ in (5.21)
and use once again the quadratic equation satisfied by $r$ to obtain
$$\left[\delta_1+4\delta_2h+\frac{\delta_1\delta_2}{\delta_0}
+4\delta_2'(1-h^2)\right]+r\left[2\delta_0+2\delta_1h-2\delta_2
+\frac{\delta_1^2}{\delta_0}+4\delta_1'(1-h^2)\right]=0.$$
Next, replacing $r$ from (5.20) and performing the needed calculations we get
$$[12\delta_0(2\omega v-v^2)(1-h^2)+D]\sqrt{D}
-[3v^2(1+h)D+2\delta_0(1-h^2)D']=0.$$
By (5.20) and (5.17) one has
$$D=(1+h)^2(9v^4+24\omega^2\delta_0\zeta), \;\;
D'=(1+h)(18v^4+24\omega^2\delta_0(\zeta-1)), \;\; \zeta=\frac{1-h}{1+h}>0.$$
We replace these values in the equation above and divide the result by
$(1+h)^3$. One obtains
$$\begin{array}{l}
[9v^4+12(2\omega^2+2\omega v-v^2)\delta_0\zeta]\sqrt{9v^4+24\omega^2\delta_0\zeta}\\
-[27v^6+12(3v^4+6\omega^2v^2-4\omega^2\delta_0)\delta_o\zeta
+48\omega^2\delta_0^2\zeta^2]=0.\end{array}$$
At the end, we introduce a new variable $z$ by
$$\sqrt{9v^4+24\omega^2\delta_0\zeta}=3(v^2+z),\quad z\in (0,\infty)$$
to obtain the equation of contact points in a final form  $P(z)=0$ where
$$\begin{array}{rl}
P(z)&=3z^3+6(3v^2-2\omega v-2\omega^2)z^2\\
&+4(v+\omega)(9v^3-18\omega v^2+4\omega^2v
+4\omega^3)z+2v^2\delta_0(3v^2-8\omega^2).\end{array}$$
Taking $P(z)=3z^3+A_2z^2+A_1z+A_0$, it is easy to verify that:

$A_2>0, A_1>0, A_0>0$ for $3v^2-8\omega^2>0$;

$A_2>0, A_1>0, A_0<0$ for  $3v^2-8\omega^2<0$, $v+\omega>0$;

$A_1<0, A_0<0$ for  $v+\omega<0$;

\noindent
(recall that $v-2\omega>0$ and $\delta_0=4v^2-4\omega v-2\omega^2>0$).
By Descartes chain rule, the equation $P(z)=0$ has no positive root
if $3v^2-8\omega^2\geq 0$ and has exactly one positive root if
$3v^2-8\omega^2<0$ and $2v^2-2\omega v-\omega^2>0$. All changes of the
variables we used throughout the proof were one-to-one, hence there is
no contact point in the first case and there is just one contact point
in the second case. As a result, in the first case $C$ does not intersect
$\Gamma_u$ and therefore $\Gamma_u\subset W_+$. In the second case,
$C$ intersects $\Gamma_u$ at a unique point corresponding to some
$\bar h_0\in (-1,1)$ so that the part of $\Gamma_u$ related to $(-1,\bar h_0)$
is in $W_+$ and the remaining part is in $W_-$. Let $\varphi_0$ be the value
corresponding to $h_0$ according to (3.11) and let $T_{max}$ be the period
of the orbit going through $(\varphi_0,0)$ in the $(\varphi,\varphi')$-plane.
Then the orbits from the period annulus having a period $T>T_{max}$ will
be stable and the remaining ones unstable. Proposition 8 is completely
proved. $\Box$.

\vspace{2ex}
\noindent
Let us recall again that in the left(right) Duffing oscillator case the
period $T$ belongs to the interval $I=(T_0,\infty)$ where
$T_0=2\pi\sqrt\frac{v}{2v-4\omega}$. Lemma 3 and Proposition 8 imply:

\vspace{2ex}
\noindent
{\bf Corollary 2.} {\it Both conditions of Assumption 1 are satisfied in the
left$($right$)$ Duffing oscillator case, provided that:}

\vspace{1ex}
(i) $3v^2-8\omega^2\geq 0$;

\vspace{1ex}
(ii) $3v^2-8\omega^2<0$, $2v^2-2\omega v-\omega^2>0$
{\it and the period $T$ is sufficiently large.}

\vspace{2ex}
\noindent
{\bf Proof of Theorem I.}  Theorem I is a direct consequence of Corollary 2
and Theorem 2, taking into account that $\varphi$ does not oscillate around
zero only in the left and right Duffing oscillator cases.  $\Box$

\vspace{2ex}
\noindent
{\bf Example 5.3: Coherent single power nonlinearities.} Let
$a(u)=\beta u^{k+1}$,
$b(u)=\gamma\beta u^k$, $k\in\N$, where $|\gamma|<1$ and $v/\beta>0$.
Or more generally, one can take $a(u)=2\omega u+ \beta u^{k+1}$,
$b(u)=2\omega+\gamma\beta u^k$, where $r=(v-2\omega)/\beta>0$.
For simplicity, assume that $c_1=0$. Then (3.5) is satisfied
with $X_0=r^{1/k}$ and Proposition 1 works. Next, equation
(3.3) implies that $\varphi$ depends on $r$ alone. Moreover, one has
$$\varphi=r^{1/k}\bar\varphi, \;\; \varphi_0=r^{1/k}\bar\varphi_0,\;\;
\varphi_1=r^{1/k}\bar\varphi_1,\;\; c_2=r^\frac{k+2}{k}\bar c_2, \;\;
U(\varphi)=r^\frac{2}{k}\bar U(\bar\varphi),$$
where $\bar\varphi$, $\bar\varphi_0$, $\bar\varphi_1$, $\bar c_2$ and $\bar U$
do not depend on $v$, $\omega$ and $\beta$. Therefore we obtain
$$\ddot d(v)=\frac12\frac{d}{dv}\int_0^T(\varphi^2+\varphi'^2)dx
=\left(\frac12\frac{d}{dv}r^{2/k}\right)\int_0^T(\bar\varphi^2+\bar\varphi'^2)dx$$
$$=\frac{r^{2/k}}{k(v-2\omega)}\int_0^T(\bar\varphi^2+\bar\varphi'^2)dx
=\frac{1}{k(v-2\omega)}\int_0^T(\varphi^2+\varphi'^2)dx. \eqno(5.22)$$
Therefore $\ddot d(v)$ takes the sign of $\beta$ (and $v-2\omega$).

\vspace{2ex}
\noindent
{\large\bf 6. The perturbed gBBM equation.}

\vspace{1ex}
\noindent
Let us denote by ${\cal C} (X,Y)$
(respectively, ${\cal B} (X, Y)$) the set of all closed (respectively, bounded)
operators $S: X \to Y$. When $X=Y$, we shall write simply ${\cal C} (X)$ and
${\cal B} (X)$.
If $S\in {\cal C} (X,Y)$, denote by ${\bf G}(S)\subset X\times Y$ its graph.
Below we choose $X=Y=L^2[0,T])$. One can define a metric
$\hat\delta$ on ${\cal C}(L^2[0,T])$ as follows: for
$R,S\in {\cal C}(L^2[0,T])$, set
$$\hat\delta(R,S)=||P_R-P_S||_{{\cal B}(L^2\times L^2)}$$
where $P_R$ and $P_S$ are the orthogonal projections on the graphs of
${\bf G}(R)$ and ${\bf G}(S)$, respectively.

The following results are well known (see Kato \cite{2}, Chapter 4,
Theorems 2.14 and 2.17):

\vspace{2ex}
\noindent
{\bf Theorem A.} {\it Take $X=Y=L^2[0,T]$ and assume that
$R,S\in {\cal C}(X,Y)$, $A\in {\cal B}(X,Y)$. Then}
$$ \hat\delta(R+A,S+A)\leq 2(1+||A||^2)\hat\delta(R,S).$$

\vspace{2ex}
\noindent
{\bf Theorem B.} {\it Assume that $S\in {\cal C}(X,Y)$ and $B$ is a
$S$-bounded operator satisfying $||Bf||\leq a||f||+b||Sf||$ with $b<1$. Then}
$$R=S+B\in {\cal C}(X,Y)\;\; and \;\;\hat\delta(R,S)\leq
\frac{\sqrt{a^2+b^2}}{1-b}.$$
Let us introduce a small perturbation in (gBBM) by taking
$a(u)=2\omega u +\beta u^k$, $k=2,3$ and $b(u)=\gamma g(u)$
in the general equation (1.1) where $\gamma$ is a small real parameter.
We will denote the travelling-wave solution
corresponding to (1.1) again by $\varphi_v$ and the one corresponding
to (gBBM) (that is, when $\gamma=0$) by $\varphi_v^0$. If $T$ is the period
$\varphi_v^0$, then by Proposition 3 for sufficiently small $\gamma$
there is a smooth function $\varphi_0(\gamma)$ such that the wave solution
$\varphi_v$ of the perturbed equation which goes through the point
$(\varphi_0(\gamma),0)$ in the $(\varphi,\varphi')$-plane will have the
same period $T$. For the related operators ${\cal H}_v$ we have respectively
$$\begin{array}{l}
{\cal H}_{v}=-v\partial_x^2+v-a'(\varphi_v)+
\gamma[g(\varphi_v)\partial_x^2+ g'(\varphi_v)\varphi'_v\partial_x
+\frac12g''(\varphi_v)\varphi'^2_v+g'(\varphi_v)\varphi''_v],\\
{\cal H}_v^0=-v\partial_x^2+v-a'(\varphi_v^0).\end{array}$$

\vspace{2ex}
\noindent
{\bf Theorem 3.} {\it If Assumption} 1(i) {\it holds for the operator
${\cal H}_v^0$, then for $\gamma$ sufficiently small it
also holds for the operator ${\cal H}_v$.}

\vspace{2ex}
\noindent
{\bf Proof.} In order to apply Theorem A, we define the operator $A$
to be a multiplication by the function $-a'(\varphi_v^0)$ and take
$S=-v\partial_x^2+v$. Then denoting $B_\gamma= {\cal H}_v-{\cal
H}_v^0$, we obtain from Theorem A the estimate
$$\hat\delta({\cal H}_v,{\cal H}_v^0)= \hat\delta(S+B_\gamma+A, S+A)\leq
2(1+||A||^2)\hat\delta(S+B_\gamma, S). \eqno(6.1)$$ As
$$B_\gamma=a'(\varphi_v^0)-a'(\varphi_v)+\gamma[-v^{-1}g(\varphi_v)S
+ g'(\varphi_v)\varphi'_v\partial_x + G]$$
where
$$G=g(\varphi_v) +\frac12g''(\varphi_v)\varphi'^2_v+g'(\varphi_v)\varphi''_v,$$
for
$f\in D(S)=H_{per}^2$, we further have
$$\begin{array}{rl}
||B_\gamma f|| &\leq \max\limits_{[0,T]}|a'(\varphi_v^0)
-a'(\varphi_v)|.||f|| \\[4mm]
& +|\gamma| [\max\limits_{[0,T]}|g(\varphi_v)/v|.||Sf|| +
\max\limits_{[0,T]} |g'(\varphi_v)\varphi'_v|.||\partial_x f|| +
\max\limits_{[0,T]} |G|.||f||].\end{array}$$

 From Plancherel identity, we have
   $$\begin{array}{rl}
   ||Sf||^2 &{\displaystyle =\int_{0}^{T}{Sf.\overline{Sf}}dx
   =T\sum\limits_{k=-\infty}^{+\infty} {|\widehat{Sf}(k)|^2}} \\[4mm]
   & {\displaystyle = v^2
 \sum\limits_{k=-\infty}^{+\infty}(1+|k|^2)^2|\overline{f}(k)|^2 \geq K(v)
 \sum\limits_{k=-\infty}^{+\infty}|k|^2|\overline{f}(k)|^2
 =K(v)||\partial_xf||^2} \end{array}$$
where $K(v)$ is a constant depending only on $v$. By using
Propositions 3 and 4, we obtain that
$\max\limits_{[0,T]}|a'(\varphi_v^0)-a'(\varphi_v)|=O(\gamma)$.
Taking into account these estimates, we come to the conclusion
that the operator $B=B_\gamma$ is $S$-bounded, with constants
$a,b$ tending to zero as $\gamma\to 0$. Applying Theorem B and
inequality (6.1), we obtain
$$\hat\delta({\cal H}_v,{\cal H}_v^0)
\leq 2(1+||A||^2)\frac{\sqrt{a^2+b^2}}{1-b}=O(\gamma).
\eqno(6.2)$$ Therefore by \cite{2} (Theorem 3.16 in Chapter 4),
the operators ${\cal H}_v$ and ${\cal H}_v^0$ have the same
spectral properties. More precisely (see also \cite{AlBoHe},
p. 363), let $U$ is the open disc bounded by $\Gamma$ and
$spec({\cal H}_v^0)\cap \overline{U}=\{\alpha_0, 0\}$, where
$\alpha_0$ is the negative eigenvalue of ${\cal H}_v^0$. Choose
circular contours $\Gamma_1$ and $\Gamma_2$ contained in $U$, such
that if $U_1$ and $U_2$ are the open discs bounded by $\Gamma_1$
and $\Gamma_2$ respectively, then  $spec({\cal H}_v^0)\cap
\overline{U}_1=\{\alpha_0\}$ and  $spec({\cal H}_v^0)\cap
\overline{U}_2=\{0\}$. From (6.2) for sufficiently small $\gamma$
we have that the $spec({\cal H}_v)\cap \overline{U}_1$ and
$spec({\cal H}_v)\cap \overline{U}_2$ consist of a single, simple
eigenvalue. Since $0$ is an eigenvalue of ${\cal H}_v$ we must
have $spec({\cal H}_v)\cap \overline{U}_2=\{0\}$, which shows that
$0$ is a simple eigenvalue of ${\cal H}_v$. Similarly we obtain
that $spec({\cal H}_v)\cap \overline{U}$ consists of a finite set
of eigenvalues of total multiplicity 2. This completes the proof
of the theorem.

\vspace{2ex}
\noindent
{\bf Proof of Theorems II and III.} Take $\omega=0$. Inequality (5.22)
implies that $\ddot{d}(v)$ is positive. When $\gamma=0$, applying respectively
Proposition 7 and Corollary 1, we conclude that ${\cal H}_v^0$ has the
needed spectral properties in both cases. By Theorem 3 the same remains
true for small $|\gamma|$ as well . Hence, Theorem 2 applies to both cases.
$\Box$

\vspace{2ex}
\noindent
{\large\bf 7. Small-amplitude travelling-wave solutions.}

\vspace{1ex} \noindent
Since the expressions in (3.10) and (4.9)
are too complex, it is difficult to use them in general. Below we
are going to consider the simpler case when the periodic waves we
study are of small amplitude. That is, $\varphi_1-\varphi_0$ is
close to zero and therefore the periodic trajectory is entirely
contained in a small neighborhood of some center $(X_0,0)\in\R^2$
given by (3.5). Below we establish that an important role in the
stability of the small-amplitude travelling-wave solutions is
played by the first isochronous (or period) constant. Let us
recall that the period function has an expansion
$$T(r)=T_0+T_2r^2+T_4r^4+T_6r^6+T_8r^8+\ldots$$
with respect to $r$, the distance between the center at $(X_0,0)$ and the
intersection point of the orbit with the $x$-axis $(\varphi_0,0)$. Then
$T_{2k}$, $k=1,2,\ldots$ is the $k$th isochronous constant. When all period
constants are zero, the center is isochronous and all orbits surrounding it
have the same period $T_0$. Since $T'(\varphi_0)=-dT/dr$, the above expansion
yields immediately:

\vspace{2ex}
\noindent
{\bf Proposition 9.} {\it The period of a small-amplitude travelling-wave solution
around a non-isochronous center is not critical.}

\vspace{2ex}
 \noindent
Let us mention, however, that the
calculation of the isochronous constants is a difficult task and
the problem of isochronicity is completely solved for only few
particular cases. Denote for short
$$p=-\frac{a''(X_0)}{3(a'(X_0)-v)},\;  q=-\frac{a'''(X_0)}{3(a'(X_0)-v)},
\; m=\frac{b'(X_0)}{v-b(X_0)},\; n=\frac{b''(X_0)}{v-b(X_0)}.\eqno(7.1)$$
As we shall see below, the first isochronous constant in our case is
expressed by
$$T_2=T_0\vartheta,\quad T_0=2\pi\sqrt\frac{v-b(X_0)}{a'(X_0)-v},
\quad\vartheta=\frac{1}{16}(15p^2+3q-m^2-6pm-2n).$$

Take a small positive $\varepsilon$ and let  $\varphi_0=X_0-\varepsilon$.
Then by (3.9) we obtain
$$\varphi_1=X_0+\varepsilon +p\varepsilon^2+p^2\varepsilon^3+O(\varepsilon^4).
\eqno(7.2)$$
The expression (7.2) as well as the formulas that follow are obtained by
long and boring asymptotic calculations which we will omit here. Thus, by using
(3.8) we come to the expression
$$U(X)=\frac{\varepsilon^2(a'(X_0)-v)(1-s^2)
[1+p(1-s)\varepsilon-\frac14(p^2(1+6s)+q(1+s^2))\varepsilon^2+\ldots]}
{(v-b(X_0))[1-ms\varepsilon-\frac12(mp(1+s)+ns^2)\varepsilon^2+\ldots]}$$
Then, by (3.7) and (3.8) we obtain
$$T=T_0[1+\vartheta\varepsilon^2 +O(\varepsilon^4)].\eqno(7.3)$$
In order to calculate the sign of the derivative $\ddot d(v)$, we
prefer to apply a more direct approach, deriving first an
alternative formula instead of (4.9). By using (4.3), we obtain
$$\ddot d(v)=\frac{d}{dv}\left(\int_0^{T/2}(\varphi^2+\varphi'^2)dx\right)=
\frac{d}{dv}\left(\int_{\varphi_0}^{\varphi_1}[s^2+U(s)]\frac{ds}{\sqrt{U(s)}}
\right).$$
In the calculation below, we denote by $\dot\varphi_0$, $\dot\varphi_1$ etc.
the derivatives with respect to the parameter $v$. As $U(s)$ vanishes
at $\varphi_0$ and $\varphi_1$, then clearly
$$\frac{d}{dv}\left(\int_{\varphi_0}^{\varphi_1}\sqrt{U(s)}ds\right)=
\int_{\varphi_0}^{\varphi_1}\frac{(\dot U(s)+\dot\varphi_0 U_{\varphi_0}(s))ds}
{2\sqrt{U(s)}}.$$
To perform differentiation in the other part of the integral, we first use
a change of the variable like (3.8) and return to the initial variable after
the calculations are made. Thus,
$$\begin{array}{l}\displaystyle
\frac{d}{dv}\left(\int_{\varphi_0}^{\varphi_1}\frac{s^2ds}{\sqrt{U(s)}}
\right)=\frac{\dot\varphi_1-\dot\varphi_0}{\varphi_1-\varphi_0}
\int_{\varphi_0}^{\varphi_1} \frac{s^2ds}{\sqrt{U(s)}}\\[5mm]
\displaystyle +\int_{\varphi_0}^{\varphi_1}
\frac{(\dot\varphi_1-\dot\varphi_0)s+\dot\varphi_0\varphi_1
-\dot\varphi_1\varphi_0}
{\varphi_1-\varphi_0} \frac{2sds}{\sqrt{U(s)}}\\[5mm]
\displaystyle-\int_{\varphi_0}^{\varphi_1}\left(\dot U(s)+
\dot\varphi_0 U_{\varphi_0}(s)+ \frac{(\dot\varphi_1-\dot\varphi_0)s
+\dot\varphi_0\varphi_1-\dot\varphi_1\varphi_0} {\varphi_1-\varphi_0}
U'(s)\right) \frac{s^2ds}{2U^{3/2}(s)}. \end{array}$$
As a result, one obtains
$$\ddot d(v)=\int_{\varphi_0}^{\varphi_1}\frac{V(s)ds}{2U^{3/2}(s)},
\eqno(7.4)$$
with
$$\begin{array}{rl}
V(s)&\displaystyle =2\frac{\dot\varphi_1-\dot\varphi_0}{\varphi_1-\varphi_0}
s^2U(s) +(U(s)-s^2)(\dot U(s)+\dot\varphi_0 U_{\varphi_0}(s))\\[3mm]
&\displaystyle+(4sU(s)-s^2U'(s))
\frac{(\dot\varphi_1-\dot\varphi_0)s+\dot\varphi_0\varphi_1-
\dot\varphi_1\varphi_0} {\varphi_1-\varphi_0}. \end{array} \eqno(7.5)$$
From (3.5), (7.1) and (7.2), for the derivatives with respect to $v$ we get
respectively
$$\begin{array}{l}
\displaystyle \dot\varphi_0=\frac{X_0}{a'(X_0)-v}-\dot\varepsilon,\\[4mm]
\displaystyle \dot\varphi_1=
\frac{X_0+(p+3p^2X_0+qX_0)\varepsilon^2+O(\varepsilon^3)}{a'(X_0)-v}
+[1+2p\varepsilon+3p^2\varepsilon^2 +O(\varepsilon^3)]\dot\varepsilon.
\end{array}\eqno(7.6)$$
Replacing (7.2) and (7.6) in the expression of $V(X)$ given by (7.5),
one obtains
$$\begin{array}{rl}
V(X)&\displaystyle = \frac{(1-s^2)X_0^2}{v-b(X_0)}\left(
5+(3p-m)X_0+\frac{a'(X_0)-v}{v-b(X_0)}\right)\varepsilon^2
+O(\varepsilon^3)\\[4mm]
&\displaystyle\left. + \frac{(1-s^2)(a'(X_0)-v)\varepsilon^2
\dot\varepsilon}{v-b(X_0)}
\right[(pX_0^2+4X_0-mX_0^2)s\\[4mm]
&\displaystyle+\left(4s^2+\frac{2(a'(X_0)-v)}{v-b(X_0)}
+X_0[p(4+8s-2s^2)+2ms^2]\right.\\[4mm]
&\displaystyle\left.\left. +X_0^2[p^2(3+2s)+pm(2s^2-2s-1)
+\frac{q}{2}(1+s^2)-(2m^2+n)s^2]\right)\varepsilon
+O(\varepsilon^2)\right].\end{array}$$
The leading term of $\dot\varepsilon$ can be determined from (7.3) after
differentiation with respect to $v$. One obtains
$$\dot\varepsilon\sim-\frac{\dot T_0}{2T_2\varepsilon}=
\frac{1}{4(a'(X_0)-v)\vartheta\varepsilon}\left(
X_0m-3pX_0-1-\frac{a'(X_0)-v}{v-b(X_0)}\right).$$
Finally, replacing in the integral (7.4), we derive the formula
$$\ddot d(v)=(a'(X_0)-v)\left[\frac{T_0X_0^2}{(a'(X_0)-v)^2}
-\frac{\dot T_0^2}{2T_2} \right]+O(\varepsilon). \eqno(7.7)$$

\vspace{1ex}
\noindent
{\bf Example 1: The BBM equation.} Let $a(u)=2\omega u+\frac32 u^2$,
$b(u)=0$. Assume for definiteness that $v>0$. The other case is considered
similarly. Using the notation introduced earlier and the expression of $X_0$,
we obtain
$$a'(X_0)-v=D^{1/2}, \;\; p=-\frac{1}{D^{1/2}},\;\; q=m=n=0, \;\;
\vartheta=\frac{15}{16D}, \;\; T_0=2\pi\frac{v^{1/2}}{D^{1/4}},$$
$$T_2=\frac{15T_0}{16D},\;\; \dot{T}_0=\frac{T_0(2\omega^2-\omega v+3c_1)}
{vD},\;\; \ddot{d}(v)=\frac{T_0}{D^{1/2}}\left[X_0^2-
\frac{8(2\omega^2-\omega v+3c_1)^2}{15v^2}\right].$$
Now, expressing $c_1$ and $X_0$ through $D$, we rewrite the expression
in the brackets of $\ddot{d}(v)$ as
$$\begin{array}{l}
\displaystyle \Lambda= X_0^2-\frac{2(D-v^2+2\omega v)^2}{15v^2}\\[4mm]
={\displaystyle -\frac{2}{15v^2}}\left[D+\sqrt\frac56 vD^{1/2}
-(1-\sqrt\frac56)v(v-2\omega)\right]\times\\[4mm]
\hspace{13mm}\times \left[D-\sqrt\frac56 vD^{1/2}-(1+\sqrt\frac56)v(v-2\omega)\right].
\end{array}$$
Next, denote by $\lambda_1$, $\lambda_2$ the roots (with respect to $D^{1/2}$)
of the first multiplier, and respectively by $\lambda_3$, $\lambda_4$ the
roots of the second multiplier.

\vspace{1ex}
\noindent
a) If $0<v\leq\frac{48}{361}(9+5\sqrt\frac56)\omega$, then both multipliers are
positive and $\Lambda<0$.

\vspace{1ex}
\noindent
b) If $\frac{48}{361}(9+5\sqrt\frac56)\omega<v\leq 2\omega$,
then the first multiplier is positive, $0\leq\lambda_4<\lambda_3$ and therefore
$\Lambda>0 \Leftrightarrow \lambda_4<D^{1/2}<\lambda_3$.

\vspace{1ex}
\noindent
c) If $\lambda>2\omega$, then $0<\lambda_1<\lambda_3$,  $\lambda_2<0$,
$\lambda_4<0$. Therefore $\Lambda>0\Leftrightarrow \lambda_1<D^{1/2}<\lambda_3$.

\vspace{1ex}
\noindent
Taking a second power of the above inequalities, it follows that
$$\ddot{d}(v)>0\;\;\Leftrightarrow\;
\begin{array}{l}
\lambda_4^2-(v-2\omega)^2<6c_1< \lambda_3^2-(v-2\omega)^2\;\;
\mbox{\rm in case b),} \\
\lambda_1^2-(v-2\omega)^2<6c_1< \lambda_3^2-(v-2\omega)^2\;\;
\mbox{\rm  in case c).}\end{array}\eqno(7.8)$$
As $\lambda_k$ are homogeneous of first degree with respect to $v,\omega$,
one can reformulate (7.8) as: $\ddot{d}(v)>0$
if and only if $(\omega/v,c_1/v^2)\in\Omega\subset\R^2$ where
$\Omega$ can be explicitly written down if  needed by using the formulas of the
quadratic roots $\lambda_k$. We are not going to do this.

\vspace{2ex}
\noindent
{\bf Example 2: The modified BBM equation.} Let $a(u)=2\omega u+\beta u^3$,
$b(u)=0$, $c_1=0$.

\vspace{1ex}
\noindent
(i) $X_0=0$ (Global center or truncated pendulum case). Then
$\ddot d(v)=\frac{4\omega^2}{3\beta v^2}T_0+O(\varepsilon)$ and
$\vartheta=\frac{3\beta}{8(v-2\omega)}\neq 0$.

\vspace{1ex}
\noindent
(ii) $X_0^2=\frac{v-2\omega}{\beta}$ (Duffing oscillator). Then
$\ddot d(v)=\frac{3v^2-8\omega^2}{6\beta v^2}T_0+O(\varepsilon)$
and $\vartheta=\frac{3\beta}{4(v-2\omega)}>0.$

\vspace{1ex}
\noindent
As seen from these examples, $\ddot d(v)$ could be negative or positive,
depending on the case.

\vspace{2ex}
\noindent
{\bf Proof of Theorems IV and V.} In Examples 1 and 2 above,
we calculated the first term of $\ddot{d}(v)$ provided that $b(u)=0$
and understood when it is positive. If one take $b(u)=\gamma g(u)$ with
$|\gamma|$ small, then $\ddot{d}(v)$ will differ from the case $b(u)=0$
by a term which is $O(\gamma)$. Therefore $\ddot{d}(v)$ will keep the sign
of its first term  provided that $|\gamma|$ and the amplitude $\varepsilon$
are small enough. This means that $\ddot{d}(v)>0$ in the domain $\Omega$
from the first example and for $3v^2-8\omega^2>0$ in the Duffing oscillator
case of the second example (recall that all solutions not oscillating around
zero belong to this case). Since the operator ${\cal H}_v^0$ related to the
above cases  satisfies the spectral properties required in Assumption 1,
the statements of Theorems IV and V follow from Theorems 2 and 3. $\Box$

\vspace{2ex}
\noindent
{\bf Acknowledgement.} This research has been partially supported by
the MESC of Bulgaria through Grant MM-1403/2004.

\end{document}